\documentclass[review]{elsarticle}

\usepackage{lineno,hyperref}
\modulolinenumbers[5]

\journal{Digital Signal Processing}

\usepackage{graphicx}
\usepackage{amsmath}
\usepackage{amssymb}
\usepackage{amsfonts}
\usepackage[caption=false,font=footnotesize,skip=0pt]{subfig}
\usepackage[font=footnotesize,skip=2pt]{caption}
\usepackage{stfloats}
\usepackage{url}
\usepackage[noend]{algorithmic}
\usepackage{algorithm}
\usepackage{setspace}
\usepackage{multirow}

\newenvironment{psmallmatrix}
  {\left(\begin{smallmatrix}}
  {\end{smallmatrix}\right)}

\begin{document}

\begin{frontmatter}

\title{Geometric Interpretation of Theoretical Bounds for RSS-based Source Localization with Uncertain Anchor Positions}



\author[mymainaddress1]{Daniel~Denkovski}
\ead{danield@feit.ukim.edu.mk}

\author[mymainaddress2]{Marko~Angjelichinoski}
\ead{maa@es.aau.dk}

\author[mymainaddress1]{Vladimir~Atanasovski}
\ead{vladimir@feit.ukim.edu.mk}

\author[mymainaddress1]{Liljana~Gavrilovska}
\ead{liljana@feit.ukim.edu.mk}


\address[mymainaddress1]{Faculty of Electrical Engineering and Information Technologies, Ss. Cyril and Methodius University in Skopje, Macedonia}
\address[mymainaddress2]{Aalborg University, Denmark}

\begin{abstract}
The Received Signal Strength based source localization can encounter severe problems originating from uncertain information about the anchor positions in practice. The anchor positions, although commonly assumed to be precisely known prior to the source localization, are usually obtained using previous estimation algorithm such as GPS. This previous estimation procedure produces anchor positions with limited accuracy that result in degradations of the source localization algorithm and topology uncertainty. We have recently addressed the problem with a joint estimation framework that jointly estimates the unknown source and uncertain anchors positions and derived the theoretical limits of the framework. This paper extends the authors previous work on the theoretical performance bounds of the joint localization framework with appropriate geometric interpretation of the overall problem exploiting the properties of semi-definiteness and symmetry of the Fisher Information Matrix and the Cram{\`e}r-Rao Lower Bound and using Information and Error Ellipses, respectively. The numerical results aim to illustrate and discuss the usefulness of the geometric interpretation. They provide in-depth insight into the geometrical properties of the joint localization problem underlining the various possibilities for practical design of efficient localization algorithms.
\end{abstract}

\begin{keyword}
Received Signal Strength, joint localization framework, Fisher Information Matrix, Cram{\`e}r-Rao Lower Bound, Information Ellipse, Error Ellipse
\end{keyword}

\end{frontmatter}

\linenumbers

\section{Introduction}

Source localization is a widespread and constantly recurring research topic in the areas of signal processing and wireless communications and networking. The location information and awareness is useful in many existing and emerging wireless networking solutions such as cellular, ad-hoc, self-organizing, context-aware and cognitive radio networks \cite{1,2}. The Received Signal Strength (RSS)-based localization techniques are of particular interest due to the radio hardware prerequisites, i.e. the inherent presence of RSS measurement features in every wireless device. This means that the utilization of RSS-based localization in practice only requires software upgrades, while offering sufficient precision and fidelity for a wide range of applications.

The generic network setup for localization comprises a set of anchors (i.e. measuring sensors) distributed over a specific area and single (or multiple) wireless transmitting source(s). In the case of RSS-based localization, the anchors, upon receiving the signal broadcasted by the source, measure the received signal power. The measurements are cooperatively combined and utilized by an RSS localization algorithm that produces an estimate of the transmitting source(s) position. 

A common and key requirement for proper operation of the localization algorithm is the knowledge of the precise positions of the anchors. In practical applications, the anchor positions are usually obtained through previous estimation procedure (such as GPS or other self-localization technique). However, the previous estimation produces erroneous anchor position estimates that cause deterioration of the source localization performance and overall network topology uncertainty. We have recently been particularly active in the area of localization with uncertain anchor positions \cite{20,21,22} and there have been also other works considering dis-calibrations and uncertainties in the anchor positions \cite{25,28}, but under a different system model (based on time and frequency difference of arrival). In \cite{20} we investigate the effect of anchor position uncertainty on source localization performance and prove that severe accuracy degradations can occur. In \cite{21} we propose a joint localization framework, named Source Position Estimation for Anchor position uncertainty Reduction--SPEAR, that aims to jointly estimate the unknown positions of the sources and reduce the uncertainty of the anchor positions. The joint localization framework uses non-Bayesian estimation formalism since it models the unknown positions of the sources and the uncertain anchors as deterministic parameters. \cite{21} further introduces a Joint Maximum Likelihood (JML) localization algorithm as a typical representative of the joint localization framework, investigates its performance in typical scenarios and shows significant source localization improvements. Furthermore, it is shown that the JML can significantly reduce the initial anchor position uncertainty. Thus, the joint localization framework can be efficiently used as powerful network topology calibration tool. In \cite{22}, we analyze the problem of source localization in presence of anchor position uncertainty and the joint localization framework theoretically. In particular, our work presented therein introduces theoretical framework for evaluation and assessment of the performance of the joint localization algorithms by deriving its fundamental lower bounds. For this purpose, \cite{22} adopts the Fisher information as a measure of the parameter information that is stored in the observations. For vector parameter estimations, the Fisher information is represented by the Fisher Information Matrix (FIM). The bounds, referred as Cram{\`e}r-Rao Lower Bounds (CRLB) are then calculated using the inverse of the FIM. Summarizing the results, \cite{22} theoretically proves that it is possible to achieve arbitrary improvements in both source localization and anchor position uncertainty reduction towards alleviating or completely eliminating the effect of the initial network topology uncertainty. The JML algorithm is shown to converge towards the derived bounds in asymptotic regions which ultimately proves that the derived bounds are correct and they can be used as a benchmark in source localization problems in presence of network topology uncertainty. Although there is vast amount of papers targeting the derivation of FIM and CRLB for transmitter localization problems \cite{25,28,26,27,29}, there is no conceptually similar research effort geometrically representing and analyzing these theoretical bounds, as the one presented in this paper.

Several authors \cite{18,19,23} have noted that the FIM and the CRLB in general, posses the properties of positive definiteness and symmetry. These properties enable suitable geometric interpretation of the FIM and CRLB. In particular, they can be both geometrically interpreted as multidimensional ellipsoids. For two-dimensional parameter estimation, such as network node localization, the ellipsoids become ellipses. In the case of the FIM, they are referred as Information Ellipses (IEs) and for the CRLB -- Error Ellipses (EEs). This paper extends our previous work in \cite{22} and focuses on the geometric interpretation of the joint localization framework and its appropriate theoretical bounds. The geometrical interpretation of the FIM provides insights into the distribution of the information about the unknown positions that is stored in the observation vector by decoupling it into orthogonal spatial dimensions. Equivalently, the geometrical interpretation of the CRLB illustrates the estimation error spread, i.e. the variance of estimation error around the true value of the estimated parameter. These insights are important since they allows deeper understanding of the localization problem. The geometrical interpretation also provides insights into the possibilities for maximizing the location information in terms of problem dimensioning, network topology design, specific node mobility patterns, channel modeling etc., considering specific limitations and conditions. This can assist the definition of algorithmic rules for optimal combining of different information components, resulting in minimization of the estimation error.

Considering the benefits that stem from the geometric interpretation of the performance bounds and related metrics, the contributions of this paper are summarized as follows. The paper introduces and discusses the geometric interpretation of the general problem of parameter estimation, as well as the problem of joint estimation, emphasizing its usage in generic RSS-based localization problems. The paper formally presents the joint localization framework and the related theoretical aspect, focuses on the structure of the FIM for arbitrary network node with unknown position, identifies different location information components (such as gains and losses due to the joint estimation) and provides its general geometric representation. To the best of the authors' knowledge, no other paper has provided similar geometric interpretation of the theoretical localization performance bounds. In this regard, the main contribution of the paper to the research community is adding a new perspective in the localization systems' and algorithms' analysis. Specifically, it introduces and employs a \textit{novel methodology} for performance analysis of planar localization systems directly via the FIM, which is notably simpler to calculate and characterize than its inverse, both analytically and numerically. The basic notion in this respect is the \emph{information ellipse} that illustrates the spatial distribution of location information. These results are very important since they provide opportunity for geometric interpretation of source localization in arbitrary settings. Finally, the paper provides detailed numerical evaluations to illustrate the dependence of the ellipses' parameters on various network parameters such as number and spatial distribution of certain and uncertain anchors, the propagation model, the effect of joint localization on performance degradation or improvement.

The paper is organized as follows. Section 2 provides the geometric interpretation in the theory of vector parameter estimations, including joint estimations, in general terms, i.e. for arbitrary parameter vector. Section 3 presents the geometric interpretation of the problem of source localization in presence of anchor position uncertainty. Section 4 focuses on further discussions based on numerical results and illustrations. Section 5 concludes the paper.

\section{Geometric interpretation of the Fisher Information Matrix}

This section provides the geometric interpretation of the Fisher Information Matrix and the Cram{\`e}r-Rao Lower Bound in general terms as Information Ellipse and Error Ellipse, respectively. It also discusses the geometric interpretation of joint estimations that result in information gains and losses also presented by Fisher Information Matrices.

\subsection{Fisher Information Matrix and Cram{\`e}r-Rao Lower Bound}

Let $\boldsymbol{\theta}\in\mathbb{R}^{m}$ denote the vector of unknown, deterministic parameters that are estimated using the observation (i.e. measurement) vector $\mathbf{x}\in\mathbb{R}^{n}$. Statistically, the observation vector can be described with the probability density function $f(\mathbf{x};\boldsymbol{\theta})$ parametrized w.r.t. $\boldsymbol{\theta}$. The likelihood function of $\boldsymbol{\theta}$ given $\mathbf{x}$ is defined as $l(\boldsymbol{\theta}|\mathbf{x})\equiv f(\mathbf{x};\boldsymbol{\theta})$. Then, the information about $\boldsymbol{\theta}$ that is stored in $\mathbf{x}$ is quantified with the Fisher Information Matrix (FIM), denoted with $\mathbf{F}(\boldsymbol{\theta})\in\mathbb{R}^{m\times m}$ and calculated using the following general formula \cite{22,23}

\begin{equation}
\setlength\abovedisplayskip{-3pt}
\mathbf{F}(\boldsymbol{\theta})=\mathbb{E}_{\mathbf{x};\boldsymbol{\theta}}\left\{(\nabla_{\boldsymbol{\theta}}\ln{f(\mathbf{x};\boldsymbol{\theta})})^T\nabla_{\boldsymbol{\theta}}\ln{f(\mathbf{x};\boldsymbol{\theta})}\right\}.
\label{eq.1}
\end{equation}

The quantify $\ln{f(\mathbf{x};\boldsymbol{\theta})}=\mathcal{L}(\boldsymbol{\theta}|\mathbf{x})$ is referred as log-likelihood function. For unbiased estimation, the covariance matrix $\mathrm{Cov}(\hat{\boldsymbol{\theta}})$ of the vector parameter estimate $\hat{\boldsymbol{\theta}}$ is bounded from below with the inverse of the FIM, i.e. the Cram{\`e}r-Rao Lower Bound (CRLB) \cite{22,23}

\begin{equation}
\mathrm{Cov}(\hat{\boldsymbol{\theta}})\succeq\mathbf{F}^{-1}(\boldsymbol{\theta})=\mathrm{CRLB}(\boldsymbol{\theta}).
\label{eq.2}
\end{equation}

In general, the FIM satisfies the following two important properties:

\begin{enumerate}
\item The FIM is positive definite, i.e. $\mathbf{F}(\boldsymbol{\theta})\succ\mathbf{0}$ which means that for any real vector $\mathbf{a}\in\mathbb{R}^m$, the following inequality holds $\mathbf{a}^T\mathbf{F}(\boldsymbol{\theta})\mathbf{a}>0$;
\item The FIM is symmetric, i.e. $\mathbf{F}^{T}(\boldsymbol{\theta})=\mathbf{F}(\boldsymbol{\theta})$.
\end{enumerate}

Considering the general properties of the FIM, the CRLB can be written as follows

\begin{equation}
\mathrm{CRLB}(\boldsymbol{\theta})=\frac{\mathbf{\Theta}(\frac{\pi}{2})\mathbf{F}(\boldsymbol{\theta})\mathbf{\Theta}^{T}(\frac{\pi}{2})}{Det\left\{\mathbf{F}(\boldsymbol{\theta})\right\}},
\label{eq.9}
\end{equation}

\noindent where $\mathbf{\Theta}(\alpha)$ is the rotation matrix for angle $\alpha$. Thus, $\mathrm{CRLB}(\boldsymbol{\theta})$ is rotated (by rotation angle $\frac{\pi}{2}$) and scaled version of $\mathbf{F}(\boldsymbol{\theta})$ and it has the same general properties since matrix rotation and multiplication are linear transformations that preserve the properties of the non-inverted matrix.

\subsection{Information and Error Ellipses}

The rest of the paper focuses on two dimensional parameter vector estimation, such as network node localization. Let $\mathbf{r}\in\mathbb{R}^2$ denote the unknown position vector of a network node. Note that, in this case $\boldsymbol{\theta}=\mathbf{r}$. Considering the general properties discussed in the previous subsection, the FIM for the case of two dimensional localization can be written as follows

\begin{equation}
\mathbf{F}(\mathbf{r})=
\begin{psmallmatrix}
F_{1}^2 && \rho F_{1}F_{2} \\
\rho F_{1}F_{2} && F_{2}^2
\end{psmallmatrix},
\label{eq.3}
\end{equation}

\noindent where $\rho\in[\;-1,1\;]$ is the correlation coefficient. Using Principal Component Analysis (PCA) \cite{23}, the FIM (\ref{eq.3}) can be represented by its eigenvalues and eigenvectors as follows

\begin{equation}
\mathbf{F}(\mathbf{r})=\mathbf{\Theta}(\alpha)
\begin{psmallmatrix}
\mu && 0 \\
0 && \eta
\end{psmallmatrix}
\mathbf{\Theta}^{T}(\alpha).
\label{eq.4}
\end{equation}

It is assumed that $\mu\geq\eta\geq0$. The eigenvector corresponding to the larger eigenvalue $\mu$ is $\mathbf{v}_{\mu}=[\;\cos{\alpha}\;\sin{\alpha}\;]^T$ and the eigenvector corresponding to the smaller eigenvalue $\eta$ is $\mathbf{v}_{\eta}=[\;-\sin{\alpha}\;\cos{\alpha}\;]^T$. These vectors are the columns of the rotation matrix $\mathbf{\Theta}(\alpha)$. The parameters $\mu,\eta$ and $\alpha$ can be easily calculated using the equality between (\ref{eq.3}) and (\ref{eq.4})

\begin{equation}
\mu=\frac{F_{1}^2+F_{2}^2}{2}+\frac{1}{2}\sqrt{(F_{1}^2-F_{2}^2)^2+4\rho^2F_{1}^2F_{2}^2},
\label{eq.6}
\end{equation}

\begin{equation}
\eta=\frac{F_{1}^2+F_{2}^2}{2}-\frac{1}{2}\sqrt{(F_{1}^2-F_{2}^2)^2+4\rho^2F_{1}^2F_{2}^2},
\label{eq.7}
\end{equation}

\begin{equation}
\alpha=\frac{1}{2}\tan^{-1}{\frac{2\rho F_1F_2}{F_1^2-F_2^2}}.
\label{eq.8}
\end{equation}

\noindent Using PCA and (\ref{eq.9}), the CRLB can be similarly represented with its eigenvalues and eigenvectors

\begin{equation}
\mathrm{CRLB}(\mathbf{r})=\mathbf{\Theta}(\alpha+\frac{\pi}{2})
\begin{psmallmatrix}
\frac{1}{\eta} && 0 \\
0 && \frac{1}{\mu}
\end{psmallmatrix}
\mathbf{\Theta}^{T}(\alpha+\frac{\pi}{2}).
\label{eq.10}
\end{equation}

The FIM $\mathbf{F}(\mathbf{r})$ can be geometrically represented with an ellipse with semi-axes lengths of $\sqrt{\mu}$ and $\sqrt{\eta}$, whose principal semi-axis (the semi-axis with length $\sqrt{\mu}$) is rotated by angle $\alpha$ in a relative coordinate system. This ellipse, referred as Information Ellipse (IE), is denoted with $\mathrm{IE}(\mu,\eta,\alpha)$ and is defined as the set of points satisfying the following inequality

\begin{equation}
\mathbf{x}^T\mathbf{F}^{-1}(\mathbf{r})\mathbf{x}\leq k,
\label{eq.12}
\end{equation}

\noindent where $k\in\mathbb{R}^{+}$. The value $k$ can be chosen in accordance with the desired confidence interval. Let $P_e$ denote the probability that the estimate of the unknown parameter vector lies within the error ellipse determined with $k$. Then, the following equation holds

\begin{equation}
k=-2\ln{(1-P_e)}.
\label{eq.13}
\end{equation}

Similarly, the CRLB can be geometrically represented with an ellipse, referred as Error Ellipse (EE), denoted with $\mathrm{EE}(\frac{1}{\eta},\frac{1}{\mu},\alpha+\frac{\pi}{2})$. Note that, $\frac{1}{\eta}\geq\frac{1}{\mu}$ since, by definition $\mu\geq\eta$.

\begin{figure*}[!ht]
\centering
\subfloat[Information Ellipse]{\includegraphics[scale=0.45]{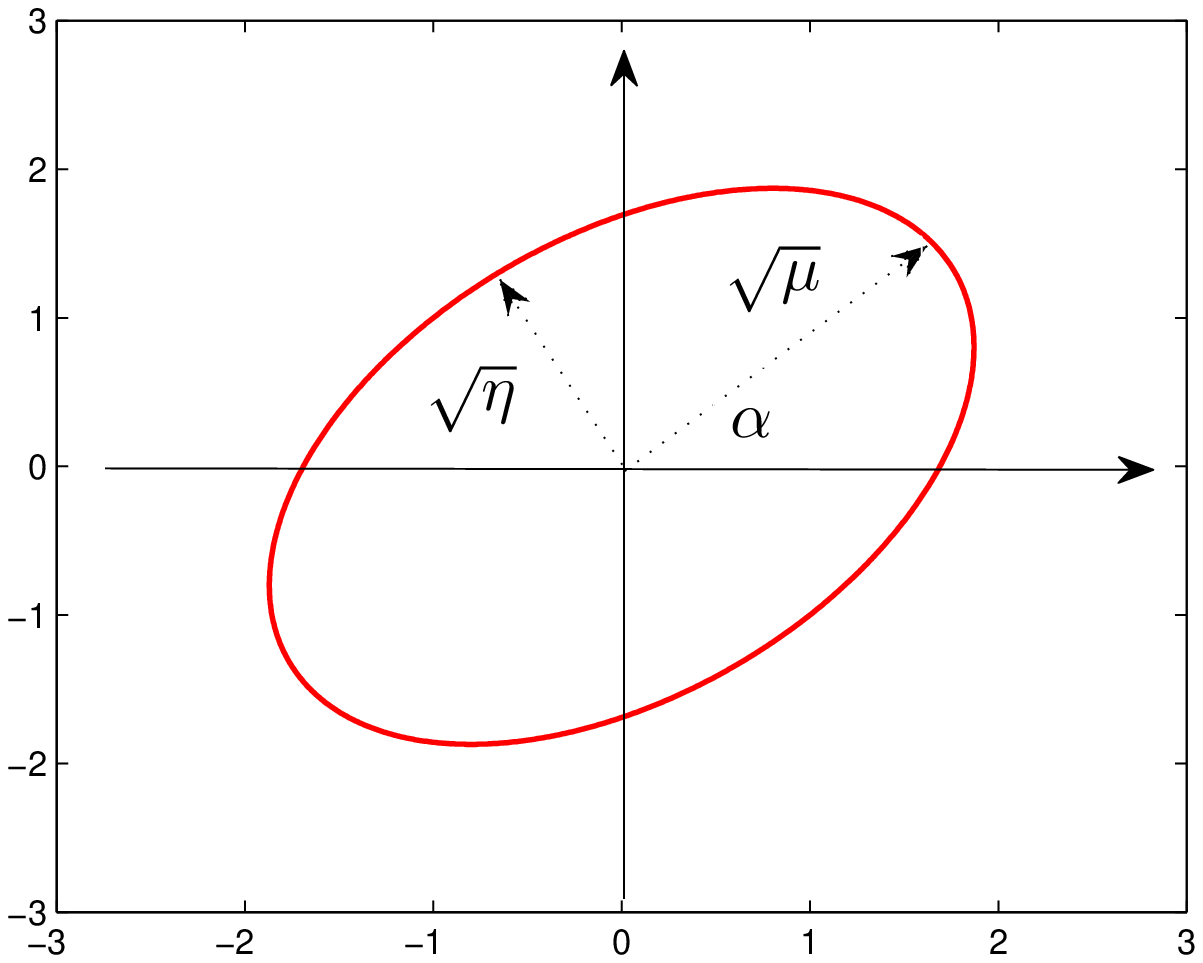}\label{0a}}
\hfil
\subfloat[Error Ellipse]{\includegraphics[scale=0.45]{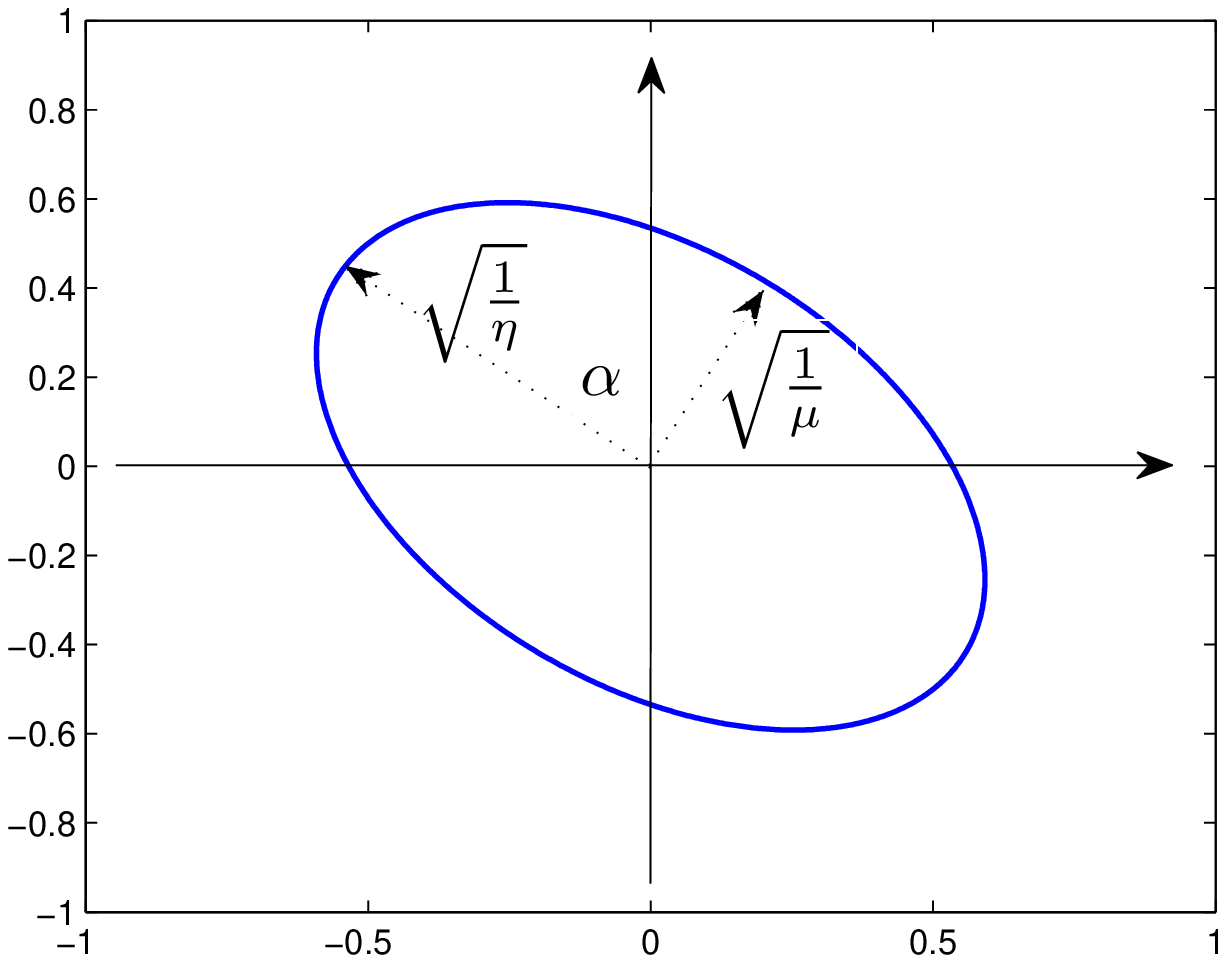}\label{0b}}
\caption{FIM and CRLB geometric interpretation}
\label{Fig0}
\end{figure*}

The IE describes the spatial distribution (the distribution in the two dimensional space spanned by the eigenvectors $\mathbf{v}_{\mu}$ and $\mathbf{v}_{\eta}$) of the information about the unknown parameter vector that is stored in the observation vector. In practical situations, it is desirable that the parameter information is distributed evenly among the dimensions. This translates to both circular and diagonal IE. As a linear measure of how close the distribution of parameter estimation is to circular, this paper uses the eccentricity of the IE defined using the following formula

\begin{equation}
\xi=\sqrt{1-\frac{\eta}{\mu}}.
\label{eq.14}
\end{equation}

Obviously, $0\leq\xi\leq1$. When $\xi\rightarrow0$, $\mu\approx\eta$ and the distribution of the parameter information among the dimensions is close to circular. However, if $\xi\rightarrow1$, $\mu>>\eta$ and the parameter information is heavily concentrated in a single dimension, which is not desirable property because the IE is close to singular. The area of an ellipse can be calculated using the standard formula

\begin{equation}
\mathrm{A}=\pi\sqrt{\mu\eta}.
\label{eq.14a}
\end{equation}

The area of the IE quantifies the amount of parameter information that is stored in the observation vector. Evidently, the EE is in indirect proportion with the area $\mathrm{A}$.

In localization problems, the Position Error Bound (PEB) is usually used to evaluate and bound the performance of localization algorithms in meters. Theoretically, the PEB is defined with $\mathcal{P}(\mathbf{r})=\sqrt{Tr\left\{\mathrm{CRLB}(\mathbf{r})\right\}}$. In terms of PCA, using (\ref{eq.10}), the PEB can be calculated using only the eigenvalues of the FIM with the following formula

\begin{equation}
\mathcal{P}(\mathbf{r})=\sqrt{\frac{1}{\mu}+\frac{1}{\eta}}=\sqrt{\frac{\mu+\eta}{\mu\eta}}.
\label{eq.11}
\end{equation}

\noindent Thus, the PEB as linear metric for localization performance evaluation and bounding is invariant in the rotation angle $\alpha$. The PEB is inversely proportional with the area of the IE, i.e.

\begin{equation}
\mathcal{P}(\mathbf{r})\sim\frac{1}{\sqrt{\mu\eta}}.
\label{eq.15}
\end{equation}

\noindent This means that the amount of parameter vector information determines the PEB for given problem setup, i.e. the information is maximized when the PEB is minimized.

\subsection{Joint estimation: Information gains and losses}

Very often, in practical signal processing problems, multiple IEs are summed as a result of adding additional parameter information, or subtracted as a result of parameter information loss. Assume that the system has on disposal two independent pieces of information about the unknown location of a node. Their respective FIMs are denoted as follows: $\mathbf{F}_1(\mathbf{r})=\mathrm{IE}(\mu_1,\eta_1,\alpha_1)$, $\mathbf{F}_2(\mathbf{r})=\mathrm{IE}(\mu_2,\eta_2,\alpha_2)$. Note that $\mu_1\geq\eta_1$ and $\mu_2\geq\eta_2$. Denote the resulting FIM with $\mathbf{F}(\mathbf{r})=\mathbf{F}_1(\mathbf{r})\pm\mathbf{F}_2(\mathbf{r})=\mathrm{IE}(\mu,\eta,\alpha)$. Note that in case of subtraction, the following condition must hold: $\mathbf{F}_1(\mathbf{r})\succ\mathbf{F}_2(\mathbf{r})$ -- otherwise, the problem is ill-conditioned. This condition translates in $\mu_1>\mu_2$ and $\eta_1>\eta_2$ as well as $\mu_1>\eta_2$ and $\eta_1>\mu_2$. It can be shown that the parameters of the new FIM can be calculated using the following important and general formulas

\begin{equation}
\mu=\frac{\mu_1+\eta_1\pm\mu_2\pm\eta_2}{2}+\frac{1}{2}\sqrt{D},
\label{eq.16}
\end{equation}

\begin{equation}
\eta=\frac{\mu_1+\eta_1\pm\mu_2\pm\eta_2}{2}-\frac{1}{2}\sqrt{D},
\label{eq.17}
\end{equation}

\begin{equation}
D=[(\mu_1-\eta_1)\pm(\mu_2-\eta_2)\cos{2(\alpha_2-\alpha_1)}]^2+(\mu_2-\eta_2)^2\sin^2{2(\alpha_2-\alpha_1)},
\label{eq.18}
\end{equation}

\begin{equation}
\alpha=\alpha_1+\frac{1}{2}\tan^{-1}{\frac{(\mu_1-\eta_1)\sin{2(\alpha_2-\alpha_1)}}{(\mu_1-\eta_1)\pm(\mu_2-\eta_2)\cos{2(\alpha_2-\alpha_1)}}}.
\label{eq.19}
\end{equation}

\noindent The updated PEB can be calculated using the following general formula

\begin{equation}
\resizebox{0.9\textwidth}{!}{$
\mathcal{P}(\mathbf{r})=\sqrt{\frac{\mu_1+\eta_1\pm\mu_2\pm\eta_2}{\mu_1\eta_1+\mu_2\eta_2\pm(\mu_1\mu_2+\eta_1\eta_2)\sin^2{(\alpha_2-\alpha_1)}\pm(\mu_1\mu_2-\eta_1\eta_2)\cos^2{(\alpha_2-\alpha_1)}}}.
$}
\label{eq.20}
\end{equation}

The results (\ref{eq.16})--(\ref{eq.19}) show that if the setup of the practical problem allows, the values of some of the network parameters can be appropriately managed such that the resulting parameter information is maximized. When summing two FIMs (i.e. two IEs), for fixed eigenvalues, the resulting node location information is maximized when the PEB is minimized, i.e. the denominator of equation (\ref{eq.20}) is maximized which is equivalent to $\alpha_2=\alpha_1\pm\frac{\pi}{2}$. Thus, the resulting ellipse $\mathrm{IE}(\mu,\eta,\alpha)$ has maximum area (for given $k$) when the parameter information update (represented with $\mathrm{IE}(\mu_2,\eta_2,\alpha_2)$) is added orthogonally to the initial parameter information (represented with $\mathrm{IE}(\mu_1,\eta_1,\alpha_1)$). In this case, the eigenvalues of the resultant FIM are $\mu_1+\eta_2$ and $\eta_1+\mu_2$ and its angle is either $\alpha_1$ if $\mu_1+\eta_2>\eta_1+\mu_2$ or $\alpha_1+\frac{\pi}{2}$ when $\mu_1+\eta_2<\eta_1+\mu_2$. Oppositely, the resulting parameter information is minimized (which is not desired in practical situations and should be avoided whenever possible) when the PEB is maximized and this is equivalent to $\alpha_2=\alpha_1$. In this case, the eigenvalues of the resulting IE are $\mu=\mu_1+\mu_2$ and $\eta=\eta_1+\eta_2$ and the angle is $\alpha=\alpha_1$.

The difference of two or more information ellipses usually arises in parameter estimation problems as a result of parameter information loss due to some form of system model uncertainty. Similarly to the case of summing two IEs, the resulting node location information is maximized when the resulting PEB is minimized or, equivalently, when the denominator in (\ref{eq.20}) is maximized. This is equivalent to $\alpha_2=\alpha_1$ which is opposite to the case when summing two IEs. The eigenvalues of the resultant IE are $\mu_1-\mu_2$ and $\eta_1-\eta_2$. Oppositely, the resulting information is minimized when the $\alpha_2=\alpha_1\pm\frac{\pi}{2}$ which produces ellipse with the following eigenvalues: $\mu_1-\eta_2$ and $\eta_1-\mu_2$.

Evidently, the distribution depends on the orientation of the different IEs (different information components), i.e. the spatial distribution of location information. The orientation of the IE, given with $\alpha$ in a relative coordinate system, gives the direction in space where the largest amount of parameter information is distributed. This corresponds to the larger semi-axis of the IE. However, the direction $\alpha+\frac{\pi}{2}$ that corresponds to the smaller semi-axis of the IE, is the direction in which the smallest amount on location information is distributed. The results show that the location information gain is largest the additional information is added in the direction of the smaller semi-axis, i.e. when adding maximum amount of information in the spatial direction with less initial information. Oppositely, the location information gain is smallest when the information update acts upon the direction of the larger semi-axis of the IE. The interpretation is opposite in the case of subtracting IEs. There, the parameter information after the reduction is maximal if the information loss component acts in the direction of the initial IE.

The results presented in this subsection and the corresponding interpretations are important in the subsequent analysis for quantifying the location information losses and gains due to anchor position uncertainties and joint estimation of the sources' and the uncertain anchors' positions.

\section{RSS-based single source localization with anchor position uncertainty}

This section formally introduces the problem of RSS-based source localization in presence of anchor position uncertainty as joint estimation problem, derives the FIM and provides its appropriate geometric interpretation using the generic theoretical approach presented in Section 2. For more intuitive interpretation of the results, the section initially presents the basic localization problem with precisely known anchor positions. This basic model is then modified to adopt and tackle anchor position uncertainty.

\subsection{Basic system model}

\begin{figure}[!ht]
\centering
\includegraphics[scale=0.45]{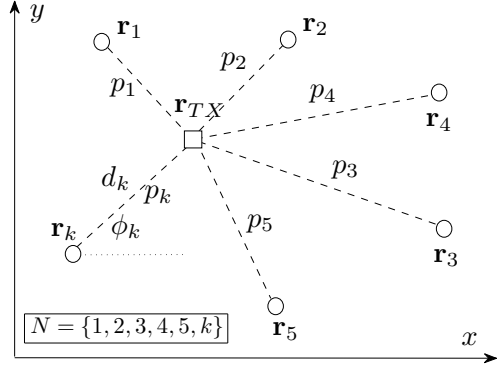}
\caption{Basic model for RSS-based single source localization}
\label{basic}
\end{figure}

The basic system model for RSS-based source localization (Fig. \ref{basic}) comprises a single wireless transmitting source whose unknown position, denoted with $\mathbf{r}_{TX}=[\;x_{TX}\;y_{TX}\;]^T$ is estimated and a set of $n$ anchors, denoted with $N=\left\{1,...,n\right\}$, arbitrarily distributed over a specific area on known positions $\mathbf{r}_k=[\;x_k\;y_k\;]^T,k\in N$. The anchors measure the strength of the received signal, broadcasted by the source. The measurements are organized in the observation vector $\mathbf{p}=[\;p_1\;...\;p_n\;]^T$, where $p_k$ denotes the RSS measurement obtained by anchor $k$. The distance between anchor $k$ and the source is denoted with $d_k=||\mathbf{r}_k-\mathbf{r}_{TX}||_2$ and the angle of the anchor $k$ in a relative coordinate system is denoted with $\phi_k=\tan^{-1}\frac{y_k-y_{TX}}{x_k-x_{TX}}$.

The goal is to obtain non-Bayesian estimator of $\boldsymbol{\theta}=\mathbf{r}_{TX}$, i.e. $\hat{\mathbf{r}}_{TX}$, that optimally exploits and combines the information stored in $\mathbf{x}$. In this context, this paper first assumes that the RSS values can be modeled using the simplified path loss model in log-normal, uncorrelated shadowing, i.e.

\begin{equation}
p_k\sim\mathcal{N}(\delta_k(\mathbf{r}_{TX}),\sigma^2),
\label{eq.30}
\end{equation}

\noindent and the mean $\delta(\mathbf{r}_{TX})$ can be written as follows

\begin{equation}
\delta(\mathbf{r}_{TX})=p_0-10\gamma\log_{10}{\big(\frac{d_k}{d_0}\big)},d_k\geq d_0.
\label{eq.31}
\end{equation}

\noindent Here, $p_0=p_{TX}-L_0$ is the received signal power at reference distance $d_0$ from the source, $p_{TX}$ and $L_0$ are the source transmit power and the reference distance propagation loss, respectively, $\gamma$ is the path loss exponent and $\sigma$ is the shadowing variance. Under this model, the log-likelihood function of $\mathbf{r}_{TX}$ is

\begin{equation}
\mathcal{L}(\mathbf{r}_{TX}|\mathbf{p})=-\frac{1}{2\sigma^2}\sum\limits_{k=1}^{n}(p_k-\delta_k(\mathbf{r}_{TX}))^2+c,
\label{eq.31.1}
\end{equation}

\noindent where $c$ is a constant independent of $\mathbf{r}_{TX}$. Using the general formula (\ref{eq.1}), the FIM for the unknown parameter $\mathbf{r}_{TX}$ can be written in the following form

\begin{equation}
\mathbf{F}(\mathbf{r}_{TX})=\sum\limits_{k=1}^n\lambda_k\mathbf{R}_k,
\label{eq.32}
\end{equation}
\vspace{3mm}
\begin{equation}
\lambda_k=\big(\frac{1}{\sigma}\frac{10\gamma}{\ln{10}}\frac{1}{d_k}\big)^2,
\label{eq.33}
\end{equation}
\vspace{3mm}
\begin{equation}
\mathbf{R}_k=
\begin{psmallmatrix}
\cos^2{\phi_k} && \cos{\phi_k}\sin{\phi_k}\\
\cos{\phi_k}\sin{\phi_k} && \sin^2{\phi_k}
\end{psmallmatrix}.
\label{eq.34}
\end{equation}

\noindent Thus, the FIM is linear combination of the matrices $\mathbf{R}_k$ for each anchor $k\in N$, with coefficients $\lambda_k$. The matrices $\mathbf{R}_k$ are symmetric ($\mathbf{R}_k=\mathbf{R}_k^T$), positive definite ($\mathbf{R}_k\succ\mathbf{0}$) and idempotent ($\mathbf{R}_k^2=\mathbf{R}_k$, which implies $\mathbf{R}_k^p=\mathbf{R}_k, p\in\mathbf{N}$). Furthermore, they are degenerate ($Det\left\{\mathbf{R}_k\right\}=0$) and ${\rm eig}(\mathbf{R}_k)=\left\{1,0\right\}$. The non-zero eigenvalue corresponds to the eigenvector $\mathbf{q}_k=[\;\cos{\phi_k}\;\sin{\phi_k}\;]^T\in\mathbb{R}^2$. Thus, the $\mathbf{R}_k$ matrices can be compactly written as $\mathbf{R}_k=\mathbf{q}_k\mathbf{q}_k^T$. Geometrically, they are represented with $\mathrm{IE}(1,0,\phi_k)$, i.e. with degenerate ellipse. Thus, the resulting IE for the unknown source location $\mathbf{r}_{TX}$ is sum of degenerate ellipses, i.e. $\mathrm{IE}(\mu,\eta,\phi)=\sum\limits_{k=1}^n\mathrm{IE}(\lambda_k,0,\phi_k)$. The parameters of the resulting IE are

\vspace{2mm}
\begin{equation}
\mu=\frac{\sum\limits_{k=1}^n\lambda_k}{2}+\frac{1}{2}\sqrt{\sum\limits_{k=1}^n\lambda_k^2+2\sum\limits_{i=1}^{n-1}\sum\limits_{j>i}\lambda_i\lambda_j\cos{2(\phi_i-\phi_j)}},
\label{eq.35.1}
\end{equation}
\vspace{7mm}
\begin{equation}
\eta=\frac{\sum\limits_{k=1}^n\lambda_k}{2}-\frac{1}{2}\sqrt{\sum\limits_{k=1}^n\lambda_k^2+2\sum\limits_{i=1}^{n-1}\sum\limits_{j>i}\lambda_i\lambda_j\cos{2(\phi_i-\phi_j)}},
\label{eq.35.2}
\end{equation}
\vspace{5mm}
\begin{equation}
\phi=\tan^{-1}{\frac{\sum\limits_{k=1}^n\lambda_k\sin{2\phi_k}}{\sum\limits_{k=1}^n\lambda_k\cos{2\phi_k}}}.
\label{eq.35.3}
\end{equation}

As an example, if the anchors are equidistantly placed on a circle with radii $d$ and the source is placed in the center of the circle, i.e. the exact position of the source, is $\mathbf{r}_{TX}=[\;0\;0\;]^T$ (in which case $d_k\equiv d,\forall k\in N$, which means that $\lambda_k\equiv\lambda,\forall k\in N$) then, the parameters of the IE of the unknown position of the transmitting source are

\begin{equation}
\mu=\frac{n\lambda}{2}+\frac{1}{2}\sqrt{n\lambda^2+2\lambda^2S},
\label{eq.36}
\end{equation}
\vspace{1mm}
\begin{equation}
\eta=\frac{n\lambda}{2}-\frac{1}{2}\sqrt{n\lambda^2+2\lambda^2S},
\label{eq.37}
\end{equation}
\vspace{2mm}
\begin{equation}
S=\sum\limits_{i=1}^{n-1}\sum\limits_{j>i}\cos{[2(\phi_i-\phi_j)]}.
\label{eq.38}
\end{equation}

\noindent The solution of the sum in (\ref{eq.38}) can be easily shown to be

\begin{equation}
S=-\frac{n}{2}.
\label{eq.39}
\end{equation}

\noindent Therefore, the IE parameters are

\begin{equation}
\mu=\eta=\frac{n\lambda}{2}.
\label{eq.40}
\end{equation}

\noindent Thus, the IE for the unknown position of the source for this geometry is circle with radii $\frac{n\lambda}{2}$.

\subsection{Basic system model: Unknown model parameters}

This subsection tackles the problem of unknown propagation model for the source position estimation in the basic system model (Fig. \ref{basic}). If only the transmit power is unknown prior to the source localization, it should be jointly estimated with the unknown location of the source, i.e. the unknown parameter vector can be written as $\boldsymbol{\theta}=[\;x_{TX}\;y_{TX}\;p_{TX}\;]^T\in\mathbb{R}^3$. The FIM for the parameter vector $\boldsymbol{\theta}$ can then, be written in the following form

\begin{equation}\label{eq.new1.4}
\mathbf{F}(\boldsymbol{\theta})=
\begin{psmallmatrix}
\sum\limits_{k=1}^n\lambda_k\cos^2{\phi_k} && \sum\limits_{k=1}^n\lambda_k\cos{\phi_k}\sin{\phi_k} && \sum\limits_{k=1}^n\frac{\sqrt{\lambda_k}}{\sigma}\cos{\phi_k} \\
\sum\limits_{k=1}^n\lambda_k\cos{\phi_k}\sin{\phi_k} && \sum\limits_{k=1}^n\lambda_k\sin^2{\phi_k} && \sum\limits_{k=1}^n\frac{\sqrt{\lambda_k}}{\sigma}\sin{\phi_k} \\
\sum\limits_{k=1}^n\frac{\sqrt{\lambda_k}}{\sigma}\cos{\phi_k} && \sum\limits_{k=1}^n\frac{\sqrt{\lambda_k}}{\sigma}\sin{\phi_k} && \frac{n}{\sigma^2}
\end{psmallmatrix}.
\end{equation}

Note that the FIM does not depend on the actual value of the transmit power. This is a major consequence of the source localization using power measurements in log-domain. The equivalent FIM for the unknown source location $\mathbf{r}_{TX}$ is

\begin{equation}\label{eq.new1.5}
\mathbf{F}(\mathbf{r}_{TX})=\mathbf{\Psi}-\bigtriangleup_{p_{TX}}\mathbf{\Psi},
\end{equation}

\noindent where

\begin{equation}\label{eq.new1.6}
\begin{split}
\bigtriangleup_{p_{TX}}\mathbf{\Psi} =\frac{1}{n}
\begin{psmallmatrix}
\sum\limits_{k=1}^n\sqrt{\lambda_k}\cos{\phi_k}\sum\limits_{k=1}^n\sqrt{\lambda_k}\cos{\phi_k} && \sum\limits_{k=1}^n\sqrt{\lambda_k}\cos{\phi_k}\sum\limits_{k=1}^n\sqrt{\lambda_k}\sin{\phi_k} \\
\sum\limits_{k=1}^n\sqrt{\lambda_k}\cos{\phi_k}\sum\limits_{k=1}^n\sqrt{\lambda_k}\sin{\phi_k} && \sum\limits_{k=1}^n\sqrt{\lambda_k}\sin{\phi_k}\sum\limits_{k=1}^n\sqrt{\lambda_k}\sin{\phi_k} 
\end{psmallmatrix} \\
 =\frac{1}{n}\mathbf{\Psi}+\frac{1}{n}
\begin{psmallmatrix}
2\sum\limits_{i=1}^{n-1}\sum\limits_{j>i}\sqrt{\lambda_i\lambda_j}\cos{\phi_i}\cos{\phi_j} && \sum\limits_{i=1}^{n-1}\sum\limits_{j>i}\sqrt{\lambda_i\lambda_j}\sin{(\phi_i+\phi_j)}\\
\sum\limits_{i=1}^{n-1}\sum\limits_{j>i}\sqrt{\lambda_i\lambda_j}\sin{(\phi_i+\phi_j)} && 2\sum\limits_{i=1}^{n-1}\sum\limits_{j>i}\sqrt{\lambda_i\lambda_j}\sin{\phi_i}\sin{\phi_j}
\end{psmallmatrix}
\end{split}
\end{equation}

\noindent The information loss matrix $\bigtriangleup_{p_{TX}}\mathbf{\Psi}$ has the following geometric representation

\begin{equation}\label{eq.new1.7}
\bigtriangleup_{p_{TX}}\mathbf{\Psi}=\mathbf{F}(a_1,0,\beta_1),
\end{equation}

\noindent with parameters 

\begin{equation}\label{eq.new1.8}
a_1=\frac{1}{n}\sum\limits_{k=1}^n\lambda_k+\frac{2}{n}\sum\limits_{i=1}^{n-1}\sum\limits_{j>i}\sqrt{\lambda_i\lambda_j}\cos{(\phi_j-\phi_j)},
\end{equation}
\vspace{3mm}
\begin{equation}\label{eq.new1.9}
\beta_1=\tan^{-1}\frac{\sum\limits_{k=1}^n\sqrt{\lambda_k}\sin{\phi_k}}{\sum\limits_{k=1}^n\sqrt{\lambda_k}\cos{\phi_k}}.
\end{equation}

\noindent Similarly, if only the path loss exponent is unknown, the the unknown parameter vector is $\boldsymbol{\theta}=[\;x_{TX}\;y_{TX}\;\gamma \;]^T\in\mathbb{R}^3$, and the corresponding FIM and the equivalent FIM for the unknown source location can be written as

\begin{equation}\label{eq.new.1.9}
\resizebox{0.9\hsize}{!}{$%
\mathbf{F}(\boldsymbol{\theta})={
\begin{psmallmatrix}
\sum\limits_{k=1}^n\lambda_k\cos^2{\phi_k} && \sum\limits_{k=1}^n\lambda_k\cos{\phi_k}\sin{\phi_k} && -\sum\limits_{k=1}^n\frac{10\sqrt{\lambda_k}}{\sigma}\cos{\phi_k}\log_{10}{\frac{d_k}{d_0}} \\
\sum\limits_{k=1}^n\lambda_k\cos{\phi_k}\sin{\phi_k} && \sum\limits_{k=1}^n\lambda_k\sin^2{\phi_k} && -\sum\limits_{k=1}^n\frac{10\sqrt{\lambda_k}}{\sigma}\sin{\phi_k}\log_{10}{\frac{d_k}{d_0}} \\
-\sum\limits_{k=1}^n\frac{10\sqrt{\lambda_k}}{\sigma}\cos{\phi_k}\log_{10}{\frac{d_k}{d_0}} && -\sum\limits_{k=1}^n\frac{10\sqrt{\lambda_k}}{\sigma}\sin{\phi_k}\log_{10}{\frac{d_k}{d_0}} && \sum\limits_{k=1}^n100\big(\log_{10}{\frac{d_k}{d_0}}\big)^2
\end{psmallmatrix}},
$%
}%
\end{equation}
\vspace{2mm}
\begin{equation}\label{eq.new.1.10}
\mathbf{F}(\mathbf{r}_{TX})=\mathbf{\Psi}-\bigtriangleup_{\gamma}\mathbf{\Psi},
\end{equation}

\noindent where the information loss due to unknown path loss exponent can be written as

\begin{equation}\label{eq.new.1.11}
\resizebox{0.9\hsize}{!}{$%
\bigtriangleup_{\gamma}\mathbf{\Psi}={\frac{\begin{psmallmatrix}
(\sum\limits_{k=1}^n\sqrt{\lambda_k}\cos{\phi_k}\log_{10}{\frac{d_k}{d_0}})^2 && (\sum\limits_{k=1}^n\sqrt{\lambda_k}\cos{\phi_k}\log_{10}{\frac{d_k}{d_0}})(\sum\limits_{k=1}^n\sqrt{\lambda_k}\sin{\phi_k}\log_{10}{\frac{d_k}{d_0}}) \\
(\sum\limits_{k=1}^n\sqrt{\lambda_k}\cos{\phi_k}\log_{10}{\frac{d_k}{d_0}})(\sum\limits_{k=1}^n\sqrt{\lambda_k}\sin{\phi_k}\log_{10}{\frac{d_k}{d_0}}) && (\sum\limits_{k=1}^n\sqrt{\lambda_k}\sin{\phi_k}\log_{10}{\frac{d_k}{d_0}})^2
\end{psmallmatrix}}{\sum\limits_{k=1}^n\log_{10}{\frac{d_k}{d_0}}}
}.
$%
}%
\end{equation}

\noindent Geometrically, it can be represented with the following degenerate IE

\vspace{-2mm}
\begin{equation}\label{eq.new.1.12}
\bigtriangleup_{\gamma}\mathbf{\Psi}=\mathbf{F}(a_2,0,\beta_2),
\end{equation}
\vspace{1mm}
\begin{equation}\label{eq.new.1.13}
a_2=\frac{\sum\limits_{k=1}^n\lambda_k(\log_{10}{\frac{d_k}{d_0}})^2+2\sum\limits_{i=1}^{n-1}\sum\limits_{j>i}\sqrt{\lambda_i\lambda_j}\cos{(\phi_i-\phi_j)}\log_{10}{\frac{d_i}{d_0}}\log_{10}{\frac{d_j}{d_0}}}{\sum\limits_{k=1}^n(\log_{10}{\frac{d_k}{d_0}})^2},
\end{equation}
\vspace{4mm}
\begin{equation}\label{eq.new.1.14}
\beta_2=\tan^{-1}{\frac{\sum\limits_{k=1}^n\sqrt{\lambda_k}\sin{\phi_k}\log_{10}{\frac{d_k}{d_0}}}{\sum\limits_{k=1}^n\sqrt{\lambda_k}\cos{\phi_k}\log_{10}{\frac{d_k}{d_0}}}}.
\end{equation}

When both, the transmit power and the path loss exponent are unknown, the unknown parameter vector is defined as $\boldsymbol{\theta}=[\;x_{TX}\;y_{TX}\;p_{TX}\;\gamma \;]^T\in\mathbb{R}^3$. However, it can be shown that the FIM for the above parameter vector is singular matrix. In other words, when the propagation model is completely unavailable, the source can be localized under the classical, non-Bayesian estimation framework. In such cases other tools should be utilized such as prior knowledge on the propagation model and this fall outside the paper's scope.

\subsection{General system model: Introducing anchor position uncertainty}

\begin{figure}[!htb]
\centering
\includegraphics[scale=0.45]{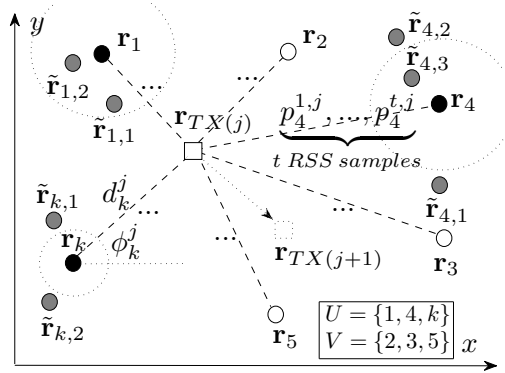}
\caption{General system model for RSS-based source localization with anchor position uncertainty}
\label{general}
\end{figure}

The general system model is shown on Fig. \ref{general}. As evident, the model is expanded to include multiple sources. The set of the sources is denoted with $S=\left\{1,...,s\right\}$ and their positions are denoted with $\mathbf{r}_{TX(j)}, j\in S$. The sources are not simultaneously active, meaning that only a single source transmits at a time. This assumption is necessary to avoid the problem of multisource localization \cite{6}. The model also assumes, that for each source position $j$, the anchor $k$ collects $t$ time RSS samples denoted with $p_{k}^{ij}, i=1,...,t$. These measurements are organized in the RSS observation vector $\mathbf{p}\in\mathbb{R}^{snt}$

\begin{equation}
\mathbf{p}=[\;\underbrace{\overbrace{p_1^{11}\;...\;p_1^{t1}\;...\;p_n^{11}\;...\;p_n^{t1}}^{nt}\;...\overbrace{\;p_1^{1s}\;...\;p_1^{ts}\;...\;p_n^{1s}\;...\;p_n^{ts}}^{nt}}_{snt}\;]^T.
\label{eq.41}
\end{equation}

The set $N$ is split into two subsets. The subset $V$ comprises all anchors whose positions are precisely known (e.g. Base Stations or fixed Access Points). The subset $U$ comprises the anchors whose positions are uncertain (e.g. they have been obtained using previous localization algorithm such as GPS) and $|U|=u$. For each uncertain anchor $k\in U$, the system has $a_k$ previous position estimates at disposal, denoted with $\tilde{\mathbf{r}}_{k,z}=[\;\tilde{x}_{k,z}\;\tilde{y}_{k,z}\;]^T, z=1,...,a_k$. These previous location estimations are organized in the following observation vector

\begin{equation}
\tilde{\mathbf{r}}=[\;\underbrace{...\;\overbrace{\tilde{\mathbf{r}}_{k,1}^T\;...\;\tilde{\mathbf{r}}_{k,a_k}^T}^{2a_k}\;...}_{2\sum\limits_{k\in U}a_k}\;]^T.
\label{eq.42}
\end{equation}

\noindent Thus, the joint observation vector $\mathbf{x}\in\mathbb{R}^{snt+2\sum\limits_{k\in U}a_k}$ in this case can be written as follows

\begin{equation}
\mathbf{x}=[\;\mathbf{x}^T\;\tilde{\mathbf{r}}^T\;]^T.
\label{eq.43}
\end{equation}

The unknown parameter vector $\boldsymbol{\theta}\in\mathbb{R}^{2s+2u}$ comprises the unknown positions of the transmitters in $S$ and the unknown position of the uncertain anchors in $U$ and can be written as

\begin{equation}
\boldsymbol{\theta}=[\;\underbrace{\mathbf{r}_{TX(1)}^T\;...\;\mathbf{r}_{TX(s)}^T}_{2s}\;\underbrace{...\;\mathbf{r}_{k}^T\;...}_{2u}\;]^T.
\label{eq.44}
\end{equation}

This paper adopts several assumptions regarding the statistical distribution of the joint observation vector $\mathbf{x}$ given with (\ref{eq.43}). Note that the assumptions are necessary to produce closed form expression for $\mathcal{L}(\boldsymbol{\theta})$, where $\boldsymbol{\theta}$ is given with (\ref{eq.44}). Initially, the paper assumes that the RSS observation vector $\mathbf{p}$ and the uncertain anchors position estimates vector $\tilde{\mathbf{r}}$ are mutually independent which results in $f(\mathbf{x};\boldsymbol{\theta})=f(\mathbf{p};\boldsymbol{\theta})f(\tilde{\mathbf{r}};\boldsymbol{\theta})$. This assumption is valid when the anchor position estimator that generates the imprecise sensor position information is independent on the currently observed RSS values, stored in $\mathbf{p}$, which can be easily achieved.

This paper also assumes spatial independence between different nodes in the network \cite{9} which translates into independence of $\mathbf{p}_k$ for each $k \in N$, and independence of $\tilde{\mathbf{r}}_k$ for each $k \in U$, as well as independence between different source positions (independence over $S$). Besides assuming spatial independence, the paper also adopts the assumption of temporal independence between the different temporal RSS samples, as well as independence between the previous position estimates for each anchor. The spatial and temporal independence assumptions are common and in many practical applications and are appropriate due to the dynamic and unpredictable nature of the wireless channel.

The previously acquired sensor position estimates are assumed to be normally distributed around the true sensor position. This assumption is natural, considering that the unbiased estimation error tends to be Gaussian in most of the cases \cite{23}. Thus

\begin{equation}
\tilde{\mathbf{r}}_k\sim\mathcal{N}(\mathbf{r}_k,\mathbf{K}_k),
\label{eq.45}
\end{equation}

\noindent where $\mathbf{K}_k$ is the covariance matrix that describes the initial anchor position uncertainty. This subsection also adopts the simplified path loss model for the distribution of the RSS samples

\begin{equation}
p_k^{ij}\sim\mathcal{N}(\delta_k^j(\boldsymbol{\theta}),\sigma^2),
\label{eq.46}
\end{equation}

\begin{equation}
\delta_k^j(\boldsymbol{\theta})=p_0-10\gamma\log_{10}\big(\frac{d_k^j}{d_0}\big); d_k^j\geq d_0.
\label{eq.47}
\end{equation}

\noindent Under all these assumptions, the log-likelihood function of $\boldsymbol{\theta}$ obtains the following form

\begin{equation}
\resizebox{0.9\hsize}{!}{$%
\mathcal{L}(\boldsymbol{\theta}|\mathbf{x})=-\frac{1}{2\sigma^2}\sum\limits_{k=1}^n\sum\limits_{j=1}^s\sum\limits_{i=1}^t(p_{k}^{ij}-\delta_{k}^{j}(\boldsymbol{\theta}))^2-\frac{1}{2}\sum\limits_{k\in U}\sum\limits_{z=1}^{a_k}(\mathbf{r}_k-\tilde{\mathbf{r}}_{k,z})^T\mathbf{K}_k^{-1}(\mathbf{r}_k-\tilde{\mathbf{r}}_{k,z})+c.
$%
}%
\label{eq.47.1}
\end{equation}

\noindent Using the result in (\ref{eq.47.1}) and the general equation in (\ref{eq.1}), the FIM $\mathbf{F}(\boldsymbol{\theta})\in\mathbb{R}^{(2s+2u)\times(2s+2u)}$ can be written in the following general form

\begin{equation}
\mathbf{F}(\boldsymbol{\theta})=
\begin{psmallmatrix}
\mathbf{\Xi} & \mathbf{\Gamma} \\
\mathbf{\Gamma}^T & \mathbf{\Omega}
\end{psmallmatrix}.
\label{eq.48}
\end{equation}

\noindent Here, the matrices $\mathbf{\Xi}$ and $\mathbf{\Omega}$ hold the location information about the unknown source positions, $\mathbf{r}_{TX(1)},...,\mathbf{r}_{TX(s)}$, and the uncertain anchors exact positions, $\mathbf{r}_k,k\in U$, respectively. Under the assumptions about spatial and temporal independence of the observations, these matrices obtain the following diagonal forms

\vspace{-2mm}
\begin{equation}
\mathbf{\Xi}=\mathrm{diag}(t\mathbf{\Psi}^1,...,t\mathbf{\Psi}^s)\in\mathbb{R}^{2s\times2s},
\label{eq.49}
\end{equation}
\vspace{-2mm}
\begin{equation}
\mathbf{\Omega}=\mathrm{diag}(...,\mathbf{\Omega}_k,...)_{k\in U}\in\mathbb{R}^{2u\times2u}.
\label{eq.50}
\end{equation}

The matrices $\mathbf{\Psi}^j, j\in S$, and $\mathbf{\Omega}_k, k\in U$, hold the information about the unknown position vectors $\mathbf{r}_{TX(j)}, j\in S$, and $\mathbf{r}_k, k\in U$, respectively, and can be written as

\begin{equation}
\mathbf{\Psi}^j=\sum\limits_{k=1}^n\lambda_k^j\mathbf{R}_k^j\in\mathbb{R}^{2\times2},
\label{eq.51}
\end{equation}
\vspace{1mm}
\begin{equation}
\mathbf{\Omega}_k=a_k\mathbf{K}_k^{-1}+t\sum\limits_{j=1}^s\lambda_k^j\mathbf{R}_k^j\in\mathbb{R}^{2\times2},
\label{eq.52}
\end{equation}

\noindent where

\begin{equation}
\lambda_k^j=\big(\frac{1}{\sigma}\frac{10\alpha}{\ln{10}}\frac{1}{d_k^j}\big)^2,
\label{eq.53}
\end{equation}
\vspace{3mm}
\begin{equation}
\mathbf{R}_k^j=
\begin{psmallmatrix}
\cos^2{\phi_k^j} & \cos{\phi_k^j}\sin{\phi_k^j} \\
\cos{\phi_k^j}\sin{\phi_k^j} & \sin^2{\phi_k^j}
\end{psmallmatrix}.
\label{eq.54}
\end{equation}

\noindent The matrix $\mathbf{\Gamma}$ couples the location information stored in both $\mathbf{\Xi}$ and $\mathbf{\Omega}$. It is calculated as

\begin{equation}
\mathbf{\Gamma}=
\begin{psmallmatrix}
t\mathbf{\Upsilon}^1 \\
\vdots \\
t\mathbf{\Upsilon}^s
\end{psmallmatrix}\in\mathbb{R}^{2s\times2u},
\label{eq.55}
\end{equation}
\vspace{3mm}
\begin{equation}
\mathbf{\Upsilon}^j=-[\:...\:\lambda_k^j\mathbf{R}_k^j\:...\:]_{k\in U}\in\mathbb{R}^{2\times2u}.
\label{eq.56}
\end{equation}

\noindent Note that the value of $\mathbf{\Gamma}$ is generally non-zero, since the vectors $\mathbf{r}_{TX(1)},...,\mathbf{r}_{TX(s)}$, and $\mathbf{r}_k,k\in U$, are not independent, i.e. they are related through the statistical model for the observation vector $\mathbf{x}$. The FIM (\ref{eq.48}) is used in the following subsections to obtain the FIM for the unknown position of arbitrary source, as well as the FIM for the unknown position of arbitrary anchor.

\subsection{The FIM for the unknown position of arbitrary source}

The FIM for the unknown position of arbitrary source $\mathbf{r}_{TX(j)},j\in S$, can be derived using the Schur's complement \cite{24}. In particular, the FIM for the unknown positions $\mathbf{r}_{TX(1)}\;...\;\mathbf{r}_{TX(s)}$ can be written as the Schur's complement of the general FIM in (\ref{eq.48})

\begin{equation}
\mathbf{F}(\mathbf{r}_{TX(1)}\;...\;\mathbf{r}_{TX(s)})=\mathbf{\Xi}-\mathbf{\Gamma}\mathbf{\Omega}^{-1}\mathbf{\Gamma}^T.
\label{eq.57.1}
\end{equation}

Using the results (\ref{eq.49})--(\ref{eq.56}), the previous result for the unknown position $\mathbf{r}_{TX(j)}$ of arbitrary source $j\in S$, can be rewritten as follows

\begin{equation}
\mathbf{F}(\mathbf{r}_{TX(1)}\;...\;\mathbf{r}_{TX(s)})=
\begin{psmallmatrix}
t\mathbf{\Psi}^{j}-t^2\mathbf{\Upsilon}^j\mathbf{\Omega}^{-1}\mathbf{\Upsilon}^{jT} && -t\mathbf{\Upsilon}^j\mathbf{\Omega}^{-1}\mathbf{\Gamma}_{S/\left\{j\right\}}^T \\
-t\mathbf{\Gamma}_{S/\left\{j\right\}}\mathbf{\Omega}^{-1}\mathbf{\Upsilon}^{jT} && \mathbf{F}(\mathbf{r}_{TX})_{S/\left\{j\right\}}
\end{psmallmatrix}.
\label{eq.57.2}
\end{equation}

\noindent Thus, the FIM for the unknown position of arbitrary source in the network can be calculated as the Schur's complement of (\ref{eq.57.2})

\begin{equation}
\resizebox{0.9\hsize}{!}{$%
\mathbf{F}(\mathbf{r}_{TX(j)})=t\mathbf{\Psi}^{j}-t^2\mathbf{\Upsilon}^j\mathbf{\Omega}^{-1}\mathbf{\Upsilon}^{jT}-t^2\mathbf{\Upsilon}^j\mathbf{\Omega}^{-1}\mathbf{\Gamma}_{S/\left\{j\right\}}^T\mathbf{F}^{-1}(\mathbf{r}_{TX})_{S/\left\{j\right\}}\mathbf{\Gamma}_{S/\left\{j\right\}}\mathbf{\Omega}^{-1}\mathbf{\Upsilon}^{jT},
$%
}%
\label{eq.57.3}
\end{equation}

\noindent or, more generally,

\begin{equation}
\mathbf{F}(\mathbf{r}_{TX(j)})=t\mathbf{\Psi}^j-\bigtriangleup\mathbf{\Psi}_1^j-\bigtriangleup\mathbf{\Psi}_2^j.
\label{eq.57}
\end{equation}

\noindent Here, $\mathbf{\Psi}^j$ is the pure information about the unknown position of the source that stems form the RSS measurements performed by the anchors. This value is multiplied by $t$ as a result of the assumption that each anchor collects $t$ independent RSS samples per source position. The quantity $\bigtriangleup\mathbf{\Psi}_1^j$, which corresponds to the second term in (\ref{eq.57.3}), is the cumulative location information loss that stems from the uncertainty of the anchors' positions in $U$ and their joint estimation with the position of the source $j$. The last quantity, $\bigtriangleup\mathbf{\Psi}_2^j$, which corresponds to the third term in (\ref{eq.57.3}), is the cumulative source location information loss that arises as a result of activating the other sources in $S/\left\{j\right\}$ and the joint estimation of their unknown positions with the source $j$ and the uncertain anchors. 

The matrix inequality $t\mathbf{\Psi}^j\succ\bigtriangleup\mathbf{\Psi}_1^j+\bigtriangleup\mathbf{\Psi}_2^j$ is always satisfied and therefore, equation (\ref{eq.57}) can be rewritten as

\begin{equation}
\mathbf{F}(\mathbf{r}_{TX(j)})=t\mathbf{\Psi}^j-\bigtriangleup\mathbf{\Psi}^j,
\label{eq.57.1.1}
\end{equation}

\noindent where $\bigtriangleup\mathbf{\Psi}=\bigtriangleup\mathbf{\Psi}_1^j+\bigtriangleup\mathbf{\Psi}_2^j$. Thus, it can be concluded that the pure source location information is always reduced as a result of the anchor position uncertainty and the estimation of their unknown positions jointly with the unknown positions of the sources. However, the magnitude or the intensity of this reduction varies and depends on the number of uncertain anchors, as well as the number of the sources on unknown positions, i.e. it varies with the structure of the sets $U$ and $S$. As an example, if the positions of the sources in $S/\left\{j\right\}$ are known, then the last term in (\ref{eq.57}) $\bigtriangleup\mathbf{\Psi}_2^j=\mathbf{0}$, or, if $U=\emptyset$ then both terms $\bigtriangleup\mathbf{\Psi}_1^j=\bigtriangleup\mathbf{\Psi}_2^j=\mathbf{0}$.

\subsection{The FIM for the unknown position of arbitrary uncertain anchor}

Similarly, the FIM for the unknown position of arbitrary uncertain anchor $\mathbf{r}_k,k\in U$, can be obtained as the Schur's complement of the general FIM in (\ref{eq.48}). In particular, the FIM for the unknown positions of the uncertain anchors $...,\mathbf{r}_k,...$, can be written in the following form

\begin{equation}
\mathbf{F}(...,\mathbf{r}_k,...)_{k\in U}=\mathbf{\Omega}-\mathbf{\Gamma}^T\mathbf{\Psi}^{-1}\mathbf{\Gamma}.
\label{eq.58.1}
\end{equation}

Using the results (\ref{eq.49})--(\ref{eq.56}), the previous result (\ref{eq.58.1}) for the unknown position of arbitrary uncertain anchor can be rewritten as follows

\begin{equation}
\mathbf{F}(...,\mathbf{r}_k,...)_{k\in U}=
\begin{psmallmatrix}
\mathbf{\Omega}_k-t\sum\limits_{j=1}^s(\lambda_k^j)^2\mathbf{R}_k^j(\mathbf{\Psi}^j)^{-1}\mathbf{R}_k^j && t\sum\limits_{j=1}^s\lambda_k^j\mathbf{R}_k^j(\mathbf{\Psi}^j)^{-1}\mathbf{\Upsilon}_{U/\left\{k\right\}}^j \\
t\sum\limits_{j=1}^s(\mathbf{\Upsilon}_{U/\left\{k\right\}}^{j})^T(\mathbf{\Psi}^j)^{-1}\lambda_k^j\mathbf{R}_k^j && \mathbf{F}(\mathbf{r}_a)_{U/\left\{k\right\}}
\end{psmallmatrix}.
\label{eq.58.2}
\end{equation}

\noindent The FIM for $\mathbf{r}_k,k\in U$, can be derived as the Schur's complement of (\ref{eq.58.2})

\begin{equation}
\mathbf{F}(\mathbf{r}_{k})_{k\in U}=\mathbf{\Omega}_k-t\sum\limits_{j=1}^s(\lambda_k^j)^2\mathbf{R}_k^j(\mathbf{\Psi}^j)^{-1}\mathbf{R}_k^j\nonumber \\
\end{equation}
\vspace{1mm}
\begin{equation}
-t^2\big(\sum\limits_{j=1}^s\lambda_k^j\mathbf{R}_k^j(\mathbf{\Psi}^j)^{-1}\mathbf{\Upsilon}_{U/\left\{k\right\}}^j\big)\mathbf{F}^{-1}(\mathbf{r}_a)_{U\left\{k\right\}}\big(\sum\limits_{j=1}^s(\mathbf{\Upsilon}_{U/\left\{k\right\}}^{j})^T(\mathbf{\Psi}^j)^{-1}\lambda_k^j\mathbf{R}_k^j\big).
\label{eq.58.3}
\end{equation}

\noindent The previous result can be rewritten in the following general form

\begin{equation}
\mathbf{F}(\mathbf{r}_k)_{k\in U}=a_k\mathbf{K}_k^{-1}+\bigtriangleup\mathbf{K}_{k,1}^{-1}-\bigtriangleup\mathbf{K}_{k,2}^{-1}-\bigtriangleup\mathbf{K}_{k,3}^{-1}
\label{eq.58}
\end{equation}

\noindent Here, $\mathbf{K}_k^{-1}$ is the initial location information about the position of the anchor $k\in U$. In general case, this quantity is multiplied by $a_k$, since the system has $a_k$ independent previous estimates of the position of the anchor on disposal. The quantity $\bigtriangleup\mathbf{K}_{k,1}^{-1}=\sum\limits_{j=1}^s\lambda_k^j\mathbf{R}_k^j$ is the main correction of the information about the position of the anchor that stems solely from the presence of $s$ transmitting sources. However, the magnitude of this correction is reduced due to several additional factors. The quantity $\bigtriangleup\mathbf{K}_{k,2}^{-1}$, corresponding to the second term in (\ref{eq.58.3}), is the location information reduction that arises when the positions of the active sources in $S$ are unknown, while the quantity $\bigtriangleup\mathbf{K}_{k,3}^{-1}$, corresponding to the third term in (\ref{eq.58.3}), is the location information loss that arises when there are other uncertain anchors in the network besides the unknown positions of the sources.

The matrix inequality $\bigtriangleup\mathbf{K}_{k,1}^{-1}\succ\bigtriangleup\mathbf{K}_{k,2}^{-1}+\bigtriangleup\mathbf{K}_{k,3}^{-1}$ is always true which means that equation (\ref{eq.58}) can be rewritten as

\begin{equation}
\mathbf{F}(\mathbf{r}_k)_{k\in U}=a_k\mathbf{K}_k^{-1}+\bigtriangleup\mathbf{K}_{k}^{-1},
\label{eq.58.1.1}
\end{equation}

\noindent where $\bigtriangleup\mathbf{K}_{k}^{-1}=\bigtriangleup\mathbf{K}_{k,1}^{-1}-(\bigtriangleup\mathbf{K}_{k,2}^{-1}+\bigtriangleup\mathbf{K}_{k,3}^{-1})\succ\mathbf{0}$. Thus, the initial anchor position information is always positively updated as a result of the joint estimation with the position of the sources. However, the magnitude of the update varies depending on the structure of the sets $S$ and $U$. As an example, if all the sources in $S$ are on known positions, then $\bigtriangleup\mathbf{K}_{k,2}^{-1}=\mathbf{0}$ and $\bigtriangleup\mathbf{K}_{k,3}^{-1}=\mathbf{0}$.

\subsection{Geometric interpretation}

The parameters of the IEs of the FIMs provided with equations (\ref{eq.57}) and (\ref{eq.58}), can be easily obtained for each specific scenario and network setup by using the results presented in Section 3. More specifically, depending on the specific scenario, the general equations (\ref{eq.57}) and (\ref{eq.58}), obtain specific forms which can be interpreted geometrically using equations (\ref{eq.35.1})--(\ref{eq.35.3}) and (\ref{eq.16})--(\ref{eq.19}). Therefore, Section 4 aims to provide illustration of the geometric interpretation of the results presented in this section, using representative network topologies and scenarios.

\section{Numerical results}

This section gives several important insights, discussions and interpretations, regarding the geometric representation of the source localization problem in presence of anchor position uncertainty, based on numerical examples in representative network topologies. Note that generalization of the obtained results over different setups is difficult, due to the tight scenario, topology and environment parameters dependency. Therefore, the presented results aim to provide general directions rather than strict conclusions.

The section uses the following representative values for the propagation model: $p_0=0dBm$, $\gamma=3.5$, $d_0=1m$ and $\sigma=5dB$. The propagation model parameters as well as the dimensions of the considered geographical area are chosen to emulate typical indoor environments. Note that the IEs and the EEs in the forthcoming subsections correspond to ellipse parameter $k=1$.

\begin{figure*}[!htb]
\begin{minipage}[!htb]{1\linewidth}
\centering
\subfloat[Circle]{\includegraphics[scale=0.25, height=1.25in]{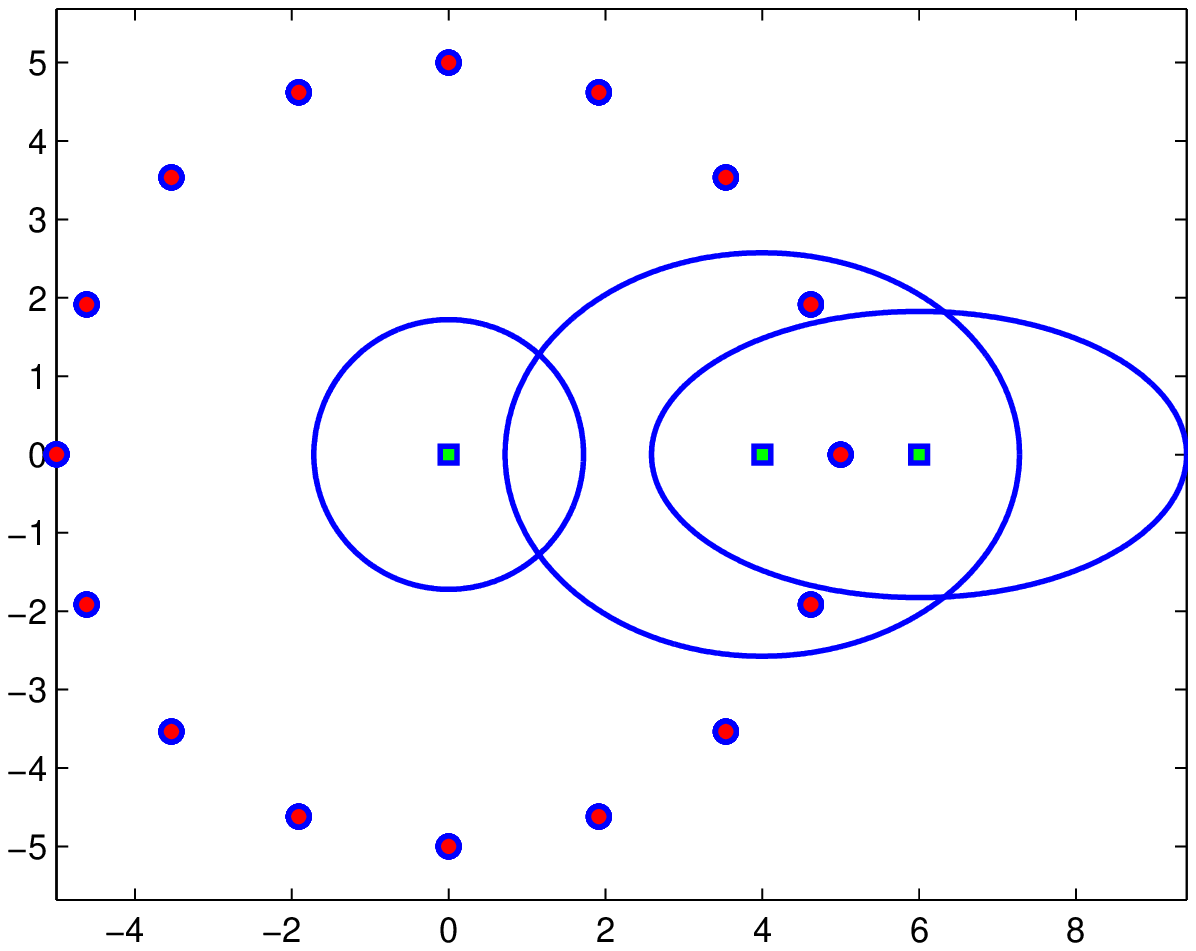}\label{1a}}
\hfil
\subfloat[Grid]{\includegraphics[scale=0.25, height=1.25in]{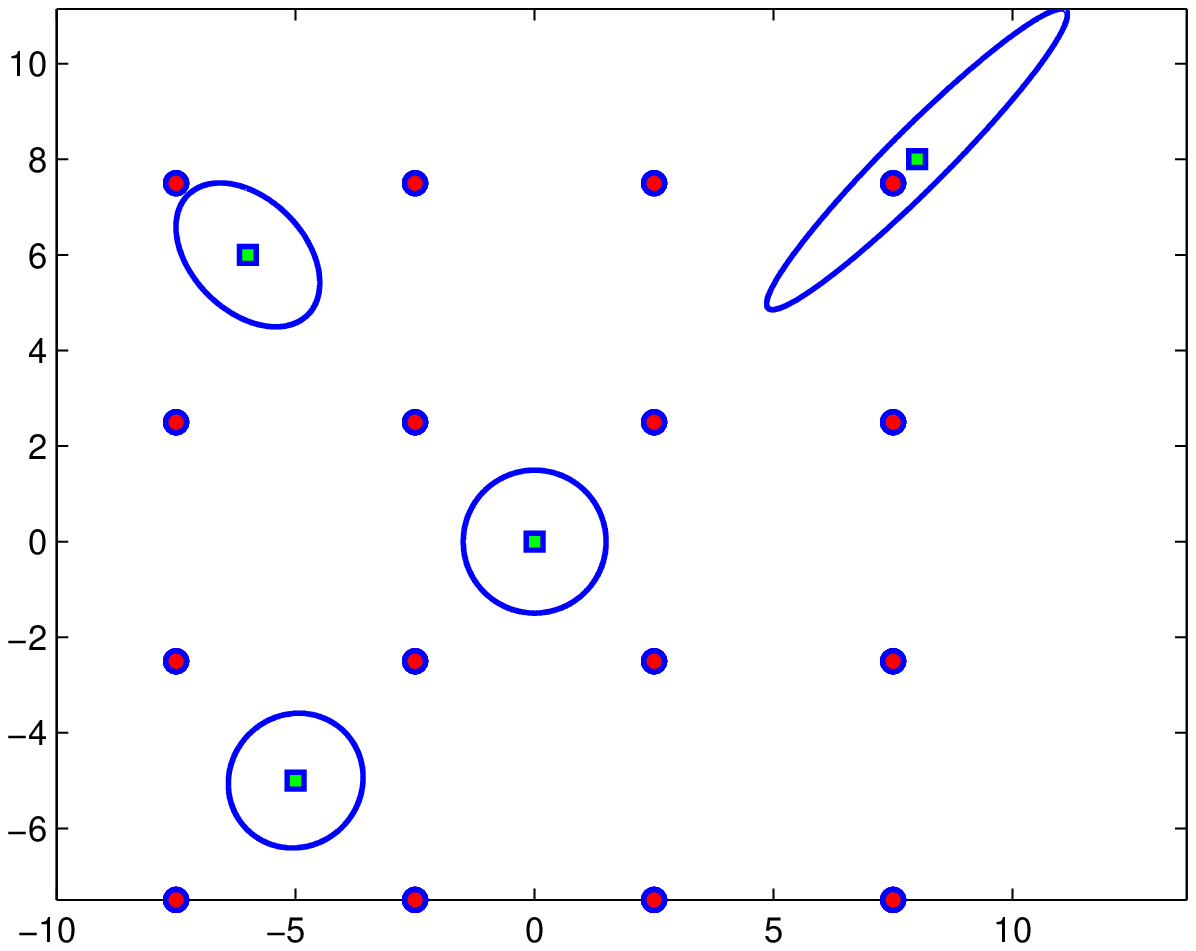}\label{1b}}
\hfil
\subfloat[Irregular distribution]{\includegraphics[scale=0.25, height=1.25in]{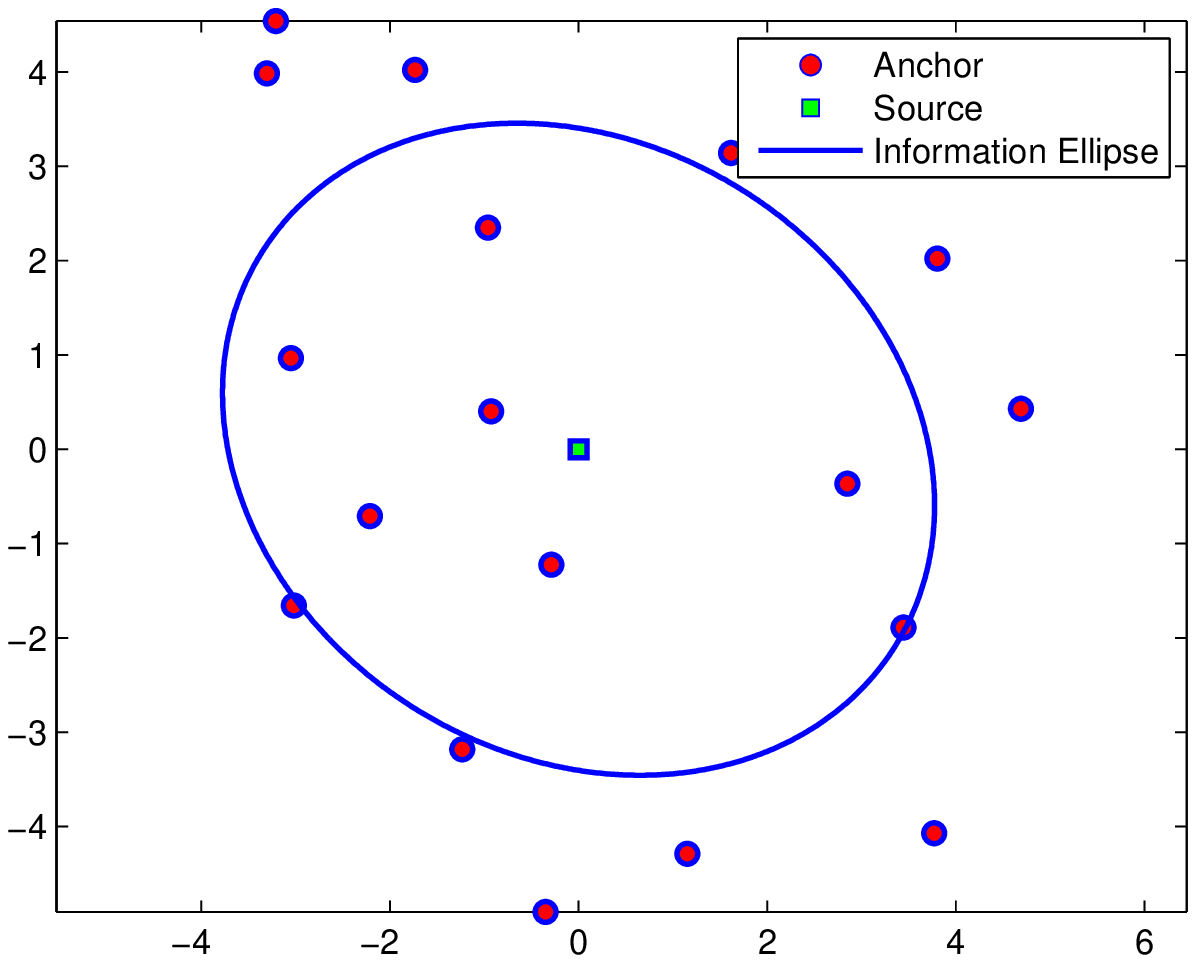}\label{1c}}
\hfil
\subfloat[Clustered distribution]{\includegraphics[scale=0.25, height=1.25in]{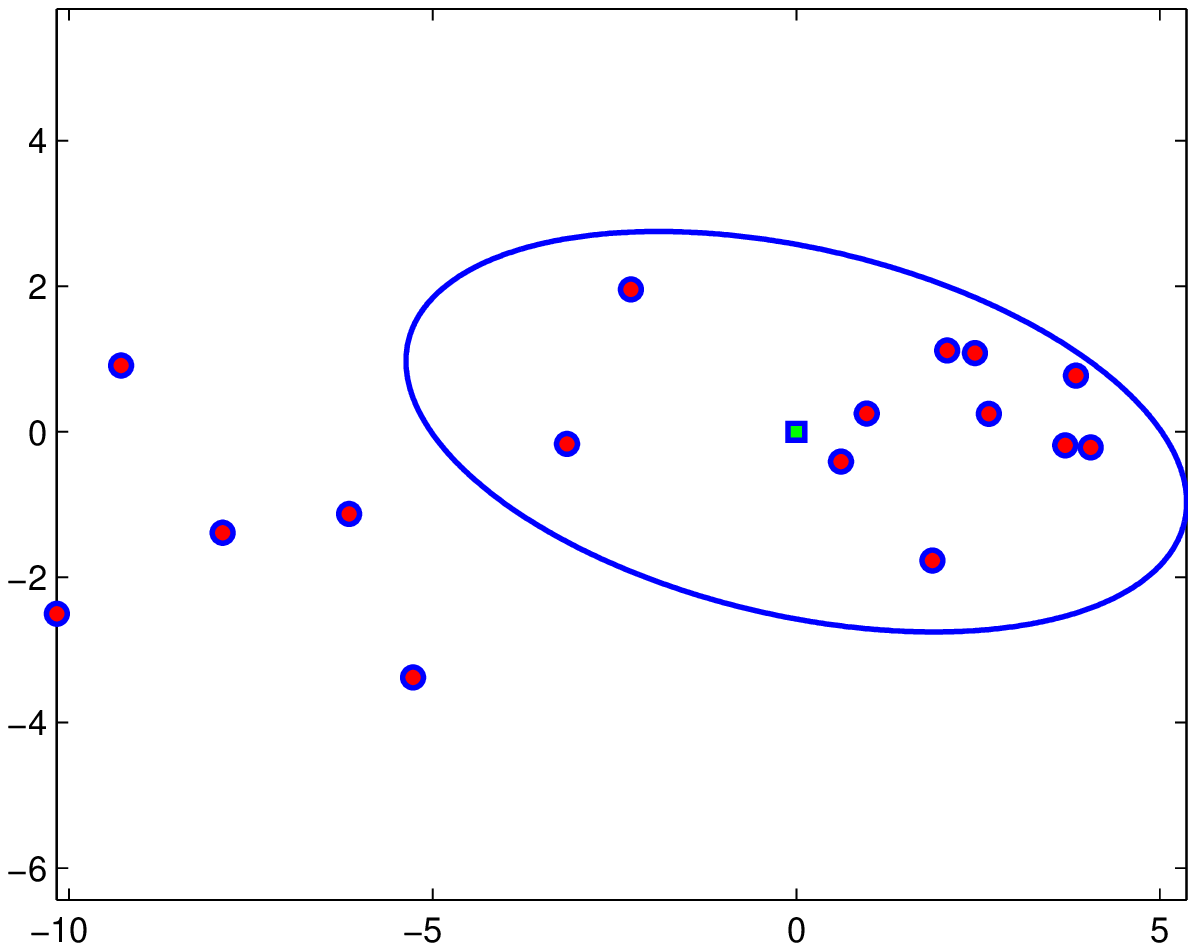}\label{1d}}
\caption{Source location IE and EE for example anchor geometries}
\vspace{5mm}
\label{Fig1}
\end{minipage}
\begin{minipage}[!htb]{.5\linewidth}
  \centering
  \includegraphics[scale=0.4, height=1.7in]{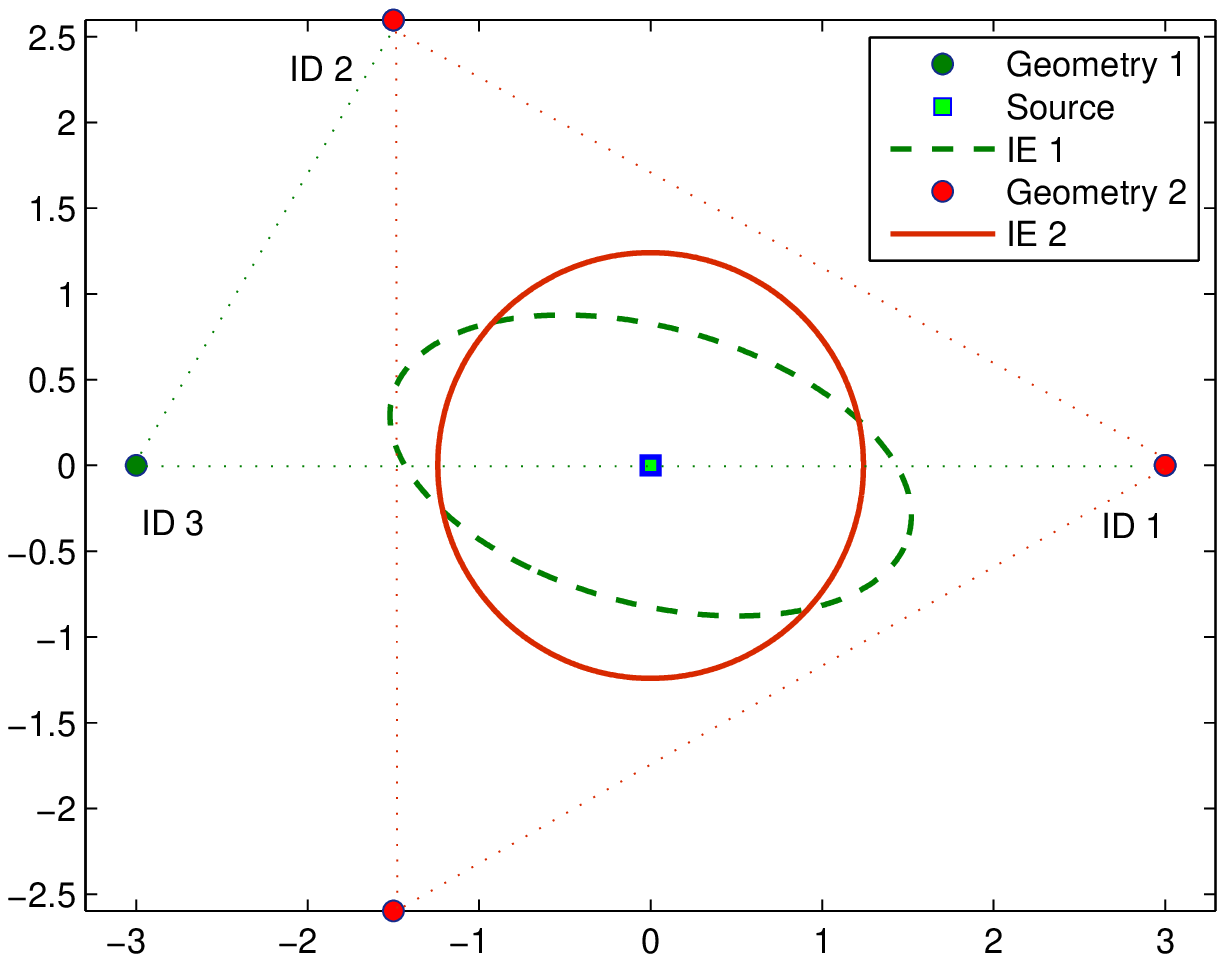}
  \captionof{figure}{The source location IE: $n=3$}
  \label{Fig2}
\end{minipage}%
\hspace{0.00mm}
\begin{minipage}[!htb]{.5\linewidth}
  \centering
  \includegraphics[scale=0.4, height=1.7in]{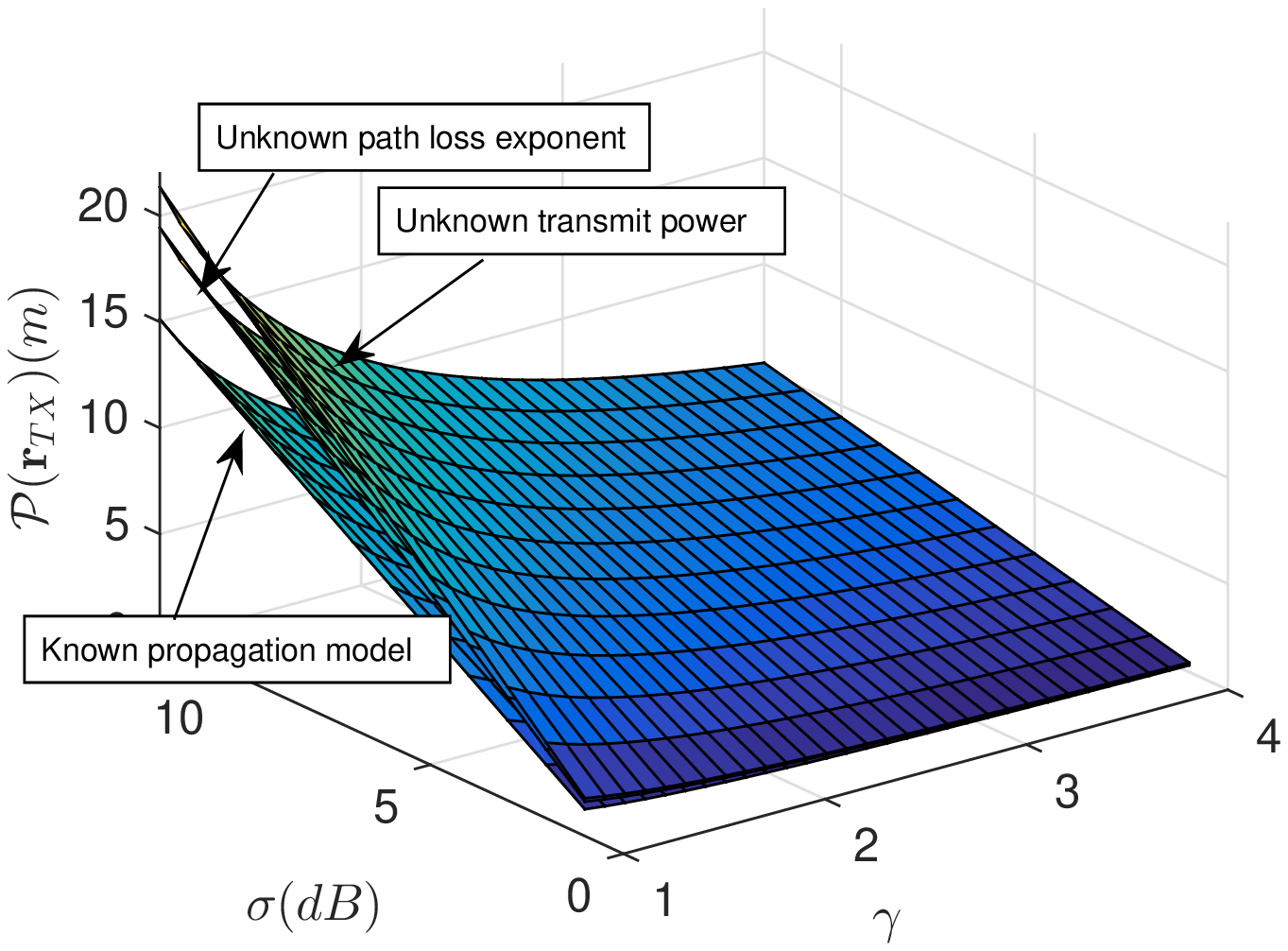}
  \captionof{figure}{Unknown model par. effect: $n=10$, $\mathbf{r}_{TX}=(0,5)$}
  \label{newFig2}
\end{minipage}
\end{figure*}

\subsection{Source location information distribution: Example anchor geometries}

This subsection illustrates the IE for the unknown source location when all anchors are with precisely known positions. Fig. \ref{Fig1} shows the IEs for various different source positions in four representative anchor geometries: circle, grid, irregular distribution and clusters. It is evident that the parameters of the IE vary.

The area of the IE increases as the number of anchors increases. This observation is obvious and can be directly derived from equations (\ref{eq.35.1}) and (\ref{eq.35.2}). However, the amount of the increase depends on the relative position of the additional anchors. Fig. \ref{Fig2} illustrates this dependence. The presented scenario shows three anchors (the minimum number of required anchors for two dimensional localization), all of them on distance $d=3m$ from the source. The anchors are organized in two different triangular geometries. In geometry 2 the anchors are equidistant, i.e. the angle between each pair is $\frac{\pi}{3}$ and in geometry 1, the angles at anchor ID:2 and anchor ID:3 are $\frac{\pi}{2}$ and $\frac{\pi}{3}$, respectively, i.e. anchors ID:1 and ID:3 are now aligned. Geometry 2 produces IE with larger area compared to geometry 1. In particular, the parameters of the resulting IE for geometry 2 are $\mu_2=\eta_2=\frac{3\lambda}{2}$ and its area is $A_2=\frac{\pi\lambda3}{2}$. The parameters of the IE for geometry 1 are $\mu_1=\frac{(3+\sqrt{3})\lambda}{2}$ and $\eta_1=\frac{(3-\sqrt{3})\lambda}{2}$ and its area is $A_1=\frac{\pi\lambda\sqrt{6}}{2}$. Evidently $A_2>A_1$. Thus, it can be concluded that for fixed number of anchors and fixed distances to the source, there is optimal anchor distribution relative to the source position that provides the optimal exploitation of the available dimensions. Evidently, in geometry 1 due to alignment, the sum source location information produced by anchors ID:1 and ID:3 is sub-optimally distributed along the line the connects these two anchors.

The larger semi-axis of the IE is always in the direction of the main source of location information. This information source can be single anchor, i.e the closest anchor (if all other anchor are relatively far away from the transmitting source compared to the closest anchor) or a set of anchors that are clustered in a specific space area. However, in anchor geometries that exhibit strong symmetry properties relative to the position of the source, as well as uniform distribution over the area, the source location information tends to a circular spatial distribution.

\begin{figure*}[!htb]
\centering
\subfloat[Eccentricity, $\phi_1=0$]{\includegraphics[scale=0.45, height=1.8in]{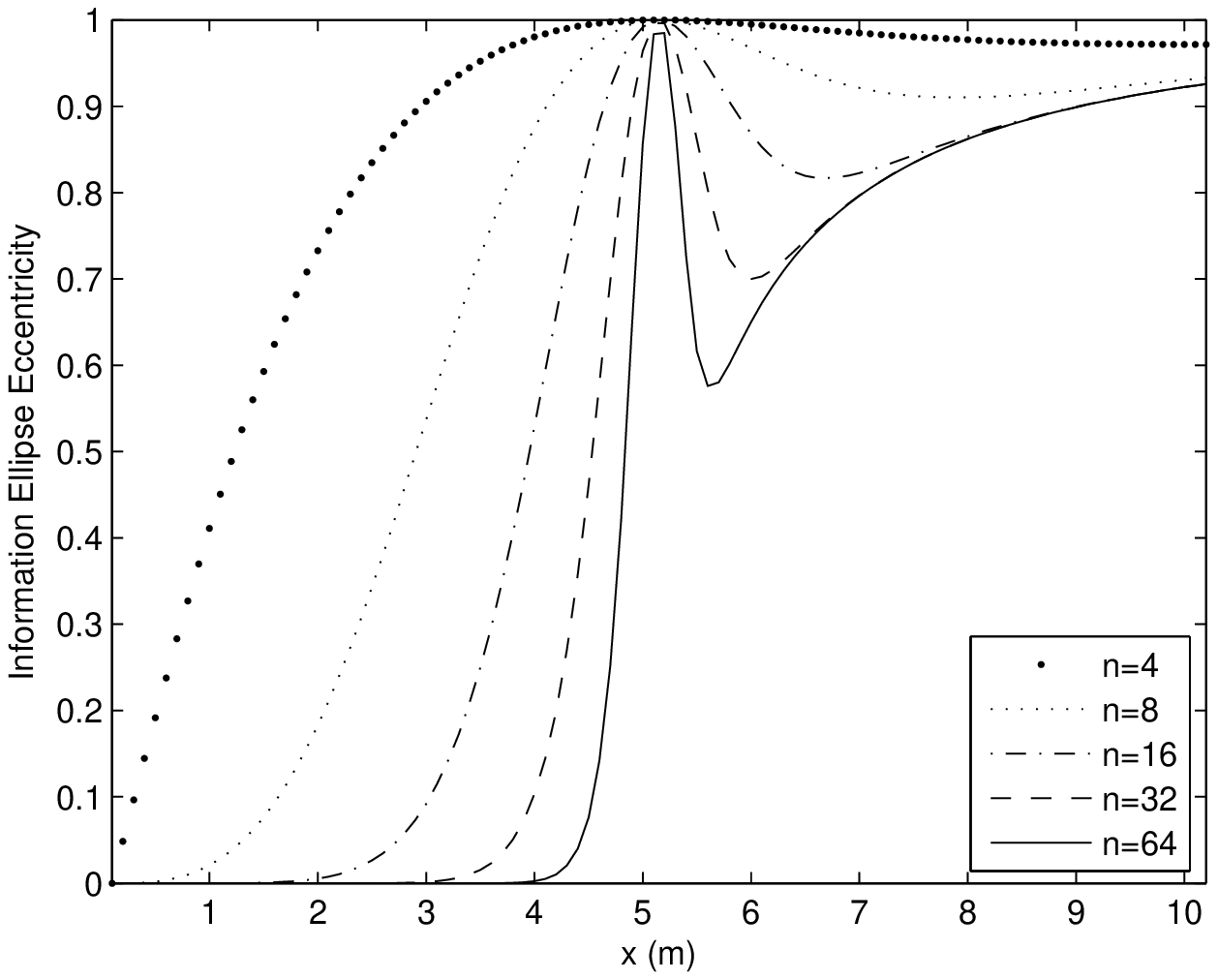}\label{4a}}
\hfil
\subfloat[Area, $\phi_1=0$]{\includegraphics[scale=0.45, height=1.8in]{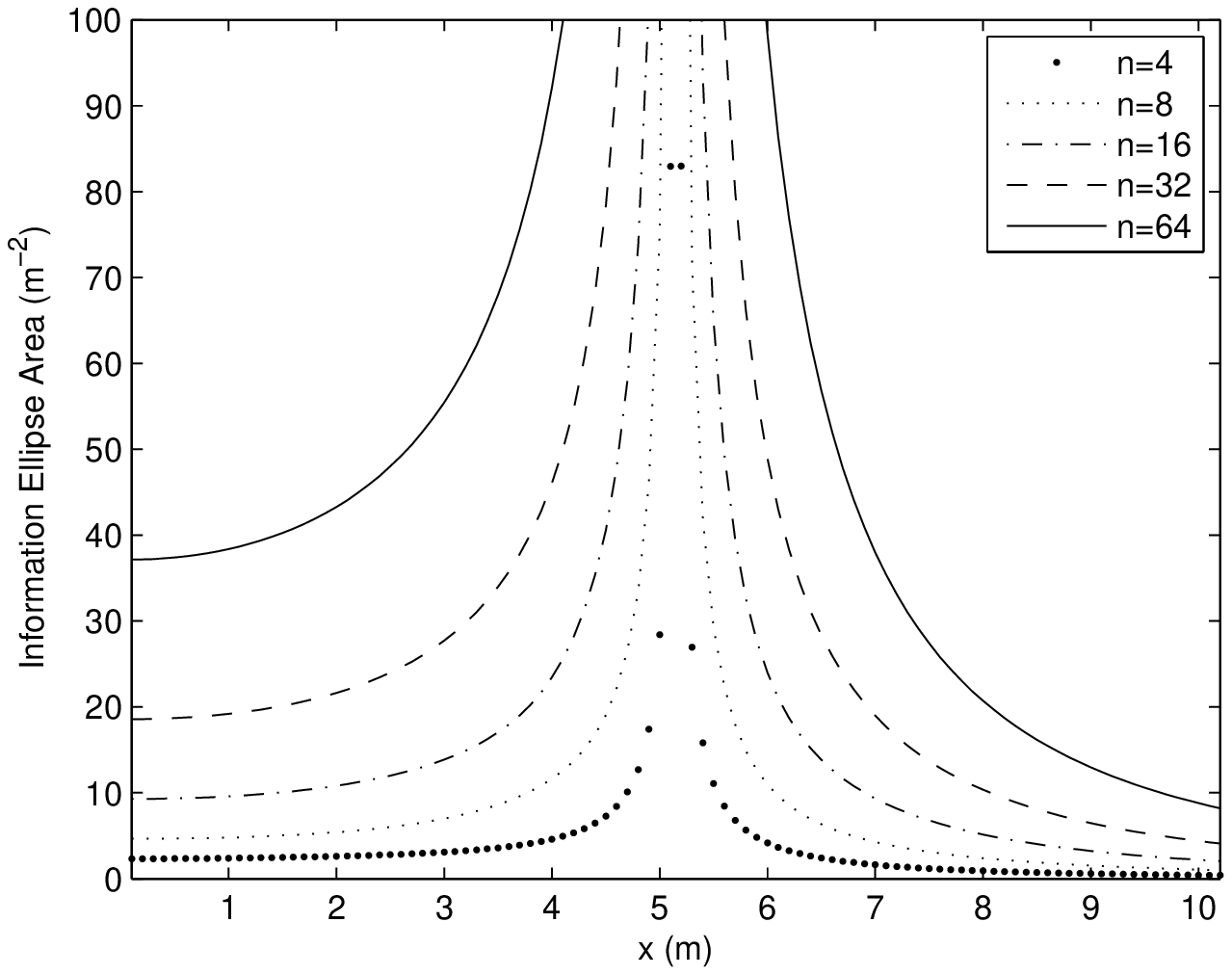}\label{4b}}
\hfil
\subfloat[Eccentricity, $\phi_1=\frac{\pi}{n}$]{\includegraphics[scale=0.45, height=1.8in]{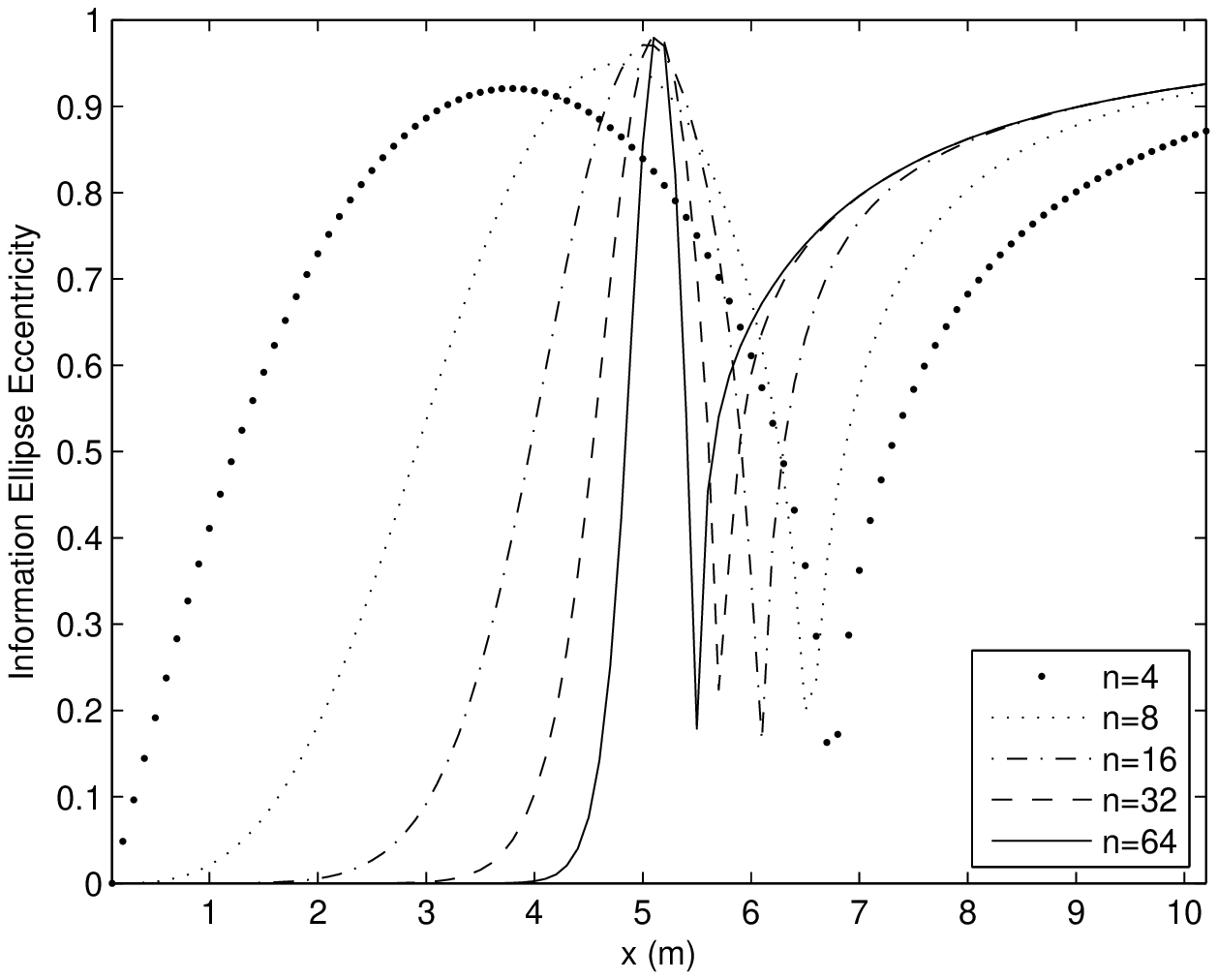}\label{4c}}
\hfil
\subfloat[Area, $\phi_1=\frac{\pi}{n}$]{\includegraphics[scale=0.45, height=1.8in]{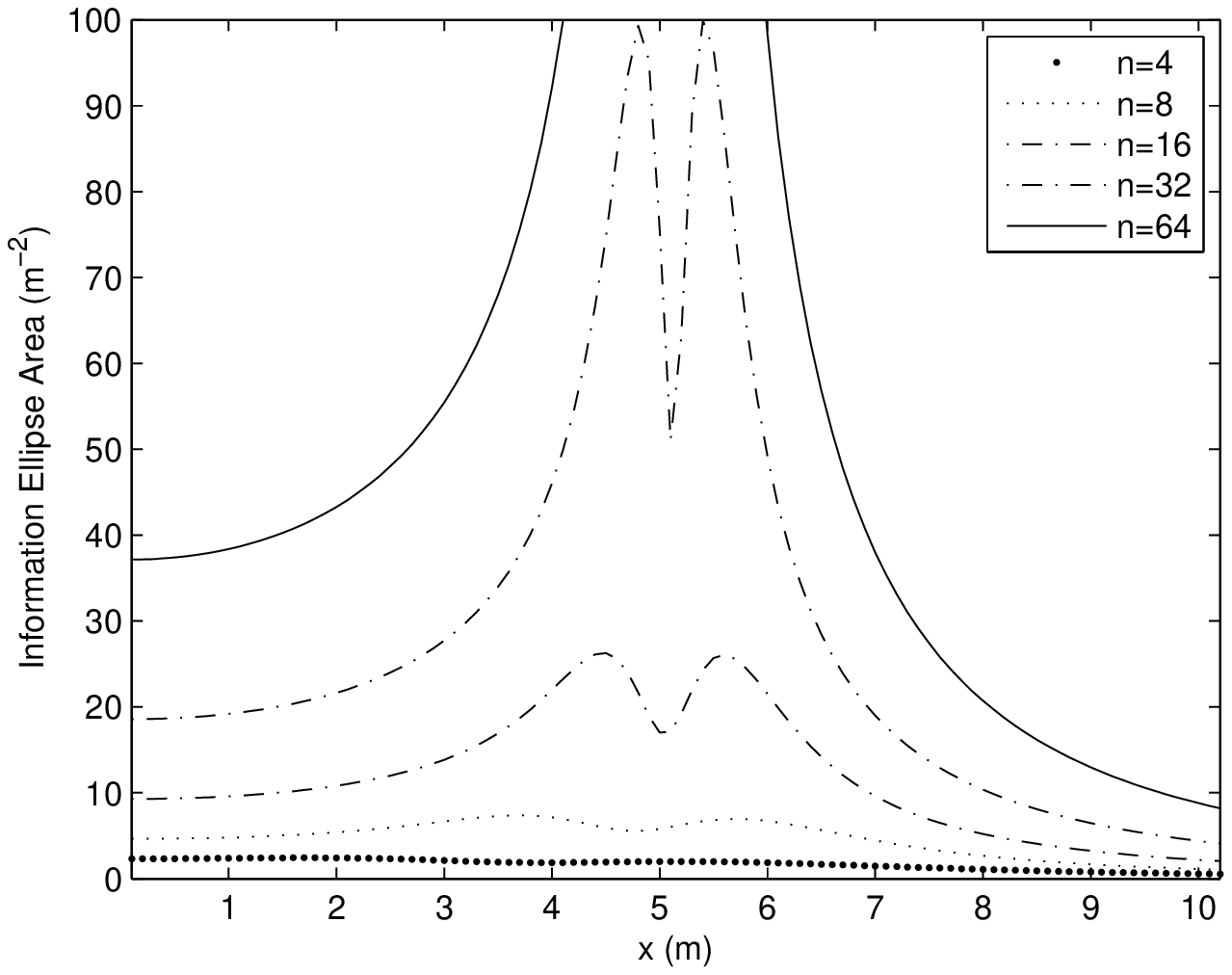}\label{4d}}
\caption{Information Ellipse eccentricity and area for two configurations of the circle anchor geometry}
\label{Fig4}
\end{figure*}

The results presented on Fig. \ref{Fig4} show the IE eccentricity and area for different source position and varying number of anchors $n$, when the anchors are equidistantly placed circle with radii $d=5m$. The angle between $i$-th anchor and the x-axis is $\phi_i=\phi_1+(i-1)\frac{2\pi}{n}, i=2,...,n$. In particular, the x-coordinate of the source position is varied in the interval $x\in[0,2d]$ while the y-coordinate remains unchanged $y=0$. The IE eccentricity and area are investigated for two different configurations of the circle anchor geometry. The results on Fig. \ref{Fig4}\subref{4a} and Fig. \ref{Fig4}\subref{4b}, are obtained for the case when $\phi_1=0$, whereas the results on Fig. \ref{Fig4}\subref{4c} and Fig. \ref{Fig4}\subref{4d}, are obtained for the case when $\phi_1=\frac{\pi}{n}$. Evidently, the IE area increases as the number of anchors increases and the eccentricity of the IE increases towards 1 as the source approaches $x=5m$, which makes the anchors in this spatial area dominant contributors towards the location information.

It can be noted that uniform spatial increase in the number of anchors results in lower IE eccentricity. As the source moves away from the anchor topology, the eccentricity approaches 1 since most of the available location information (which decreases as the area graphs show) remains distributed along the x-axis. This is due to the fact that in such case, the source observes the anchors as a single cluster in the space.

The main difference between the result presented on Fig. \ref{Fig4}\subref{4a}, Fig. \ref{Fig4}\subref{4b} and Fig. \ref{Fig4}\subref{4c}, Fig. \ref{Fig4}\subref{4d}, are the IE eccentricity and area profiles around $x=5m$. For the first anchor topology configuration when $\phi_1=0$, in the spatial area around $x=5m$, the source is very close to a single anchor, i.e. anchor 1. According to the adopted propagation model, the coefficient $\lambda_1$ becomes excessively large. This results in very steep and unbounded rise of the IE area around this point. However, as presented on Fig. 5b, when the anchor topology is configured with $\phi_1=\frac{\pi}{n}$, the total IE area exhibits small value decrease around the point $x=d\cos{(\frac{\pi}{n})}$ since here the anchors 1 and $n$ become aligned or very close to being aligned. For larger $n$, the anchors $2, 3,...$ and $n-1, n-2,...$, are also very close to being aligned in the spatial area around $x=d\cos{(\frac{\pi}{n})}$. In such case, as previously discussed, the sum location information that stems from the respective anchors is sub-optimally distributed mostly in a single dimension resulting in smaller overall IE area (as observed on Fig. \ref{Fig2}).

It can be concluded that the distribution of the source location information in the space is indeed heavily dependent on the specific scenario and setup. However, note that equations (\ref{eq.35.1}), (\ref{eq.35.2}) and (\ref{eq.35.3}) provide closed form expressions for the IE parameters which can be used in arbitrary scenario for localization algorithm performance predictions, as well as scenario dimensioning to obtain optimal performance. The following subsection illustrates the effect of anchor position uncertainty on the overall location information distribution.

Finally, Fig \ref{newFig2} illustrates the effect of unknown transmit power and path loss exponent on the source localization. It depicts the Position Error Bound (PEB) as a function of the shadowing variance and the true value of the path loss exponent, for a scenario with $n=10$ anchors, equidistantly distributed on a circle with radii $d=10m$ and a position of the source $\mathbf{r}_{TX}=(0,5)$. It is immediately evident that the source localization improves significantly as the path loss exponent increases. Moreover, in this scenario, the unknown transmit power induces larger performance loss as compared to the loss induced when the path loss exponent in unknown.

\subsection{Localization with anchor position uncertainty: the effect of information loss}

This subsection also focuses on the circular anchor distribution with fixed number of anchors $n=64$, equidistantly placed on a circle with radii $d=5m$ and $\phi_1=0$. The goal is to investigate the effect of anchor position uncertainty on the location information IE. The number of uncertain anchors is $u=16$ and they are the anchors in the first quadrant. Their uncertainty follows circular distribution, i.e. $\mathbf{K}_k=\Delta^2\mathbf{I}_2, \forall k\in U$. In this case, the FIM for the unknown position of the source can be simply written as

\begin{equation}
\mathbf{F}(\mathbf{r}_{TX})=\mathbf{\Psi}-\bigtriangleup\mathbf{\Psi},
\end{equation}

\noindent where $\bigtriangleup\mathbf{\Psi}=\sum\limits_{k\in U}\bigtriangleup\lambda_k\mathbf{R}_k$ and

\begin{equation}
\bigtriangleup\lambda_k=\lambda_k^2\Delta^2(1-\Delta^2\lambda_k(1+\lambda_k\Delta^2)^{-1})<\lambda_k.
\label{eq.new1}
\end{equation}

The IE parameters of the source position information loss, $\bigtriangleup\mathrm{IE}(\bigtriangleup\mu,\bigtriangleup\eta,\bigtriangleup\alpha)$ can be easily calculated using equation (\ref{eq.35.1}), (\ref{eq.35.2}) and (\ref{eq.35.3}), since this loss is of the same form as the pure source position information $\mathbf{\Psi}=\sum\limits_{k=1}^n\lambda_k\mathbf{R}_k$. The parameters of the resultant IE $\mathrm{IE^*}(\mu^*,\eta^*,\alpha^*)=\mathrm{IE}(\mu,\eta,\alpha)-\bigtriangleup\mathrm{IE}(\bigtriangleup\mu,\bigtriangleup\eta,\bigtriangleup\alpha)$ can be also easily calculated using the results (\ref{eq.16}), (\ref{eq.17}) and (\ref{eq.19}).

\begin{figure*}[!htb]
\begin{minipage}[!htb]{1\linewidth}
\centering
\subfloat[$x=4m$]{\includegraphics[scale=0.4, height=1.7in]{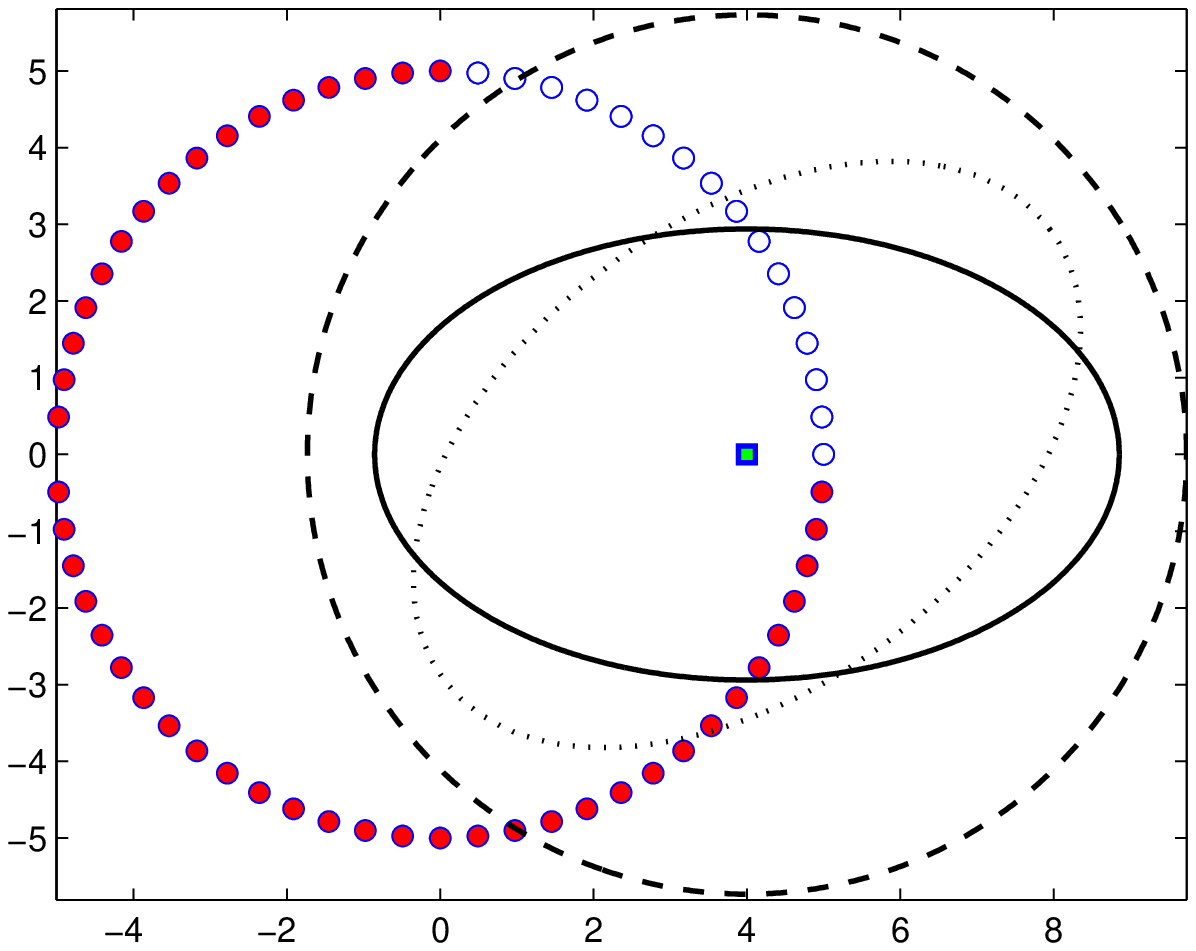}\label{6a}}
\hfil
\subfloat[$x=8m$]{\includegraphics[scale=0.4, height=1.7in]{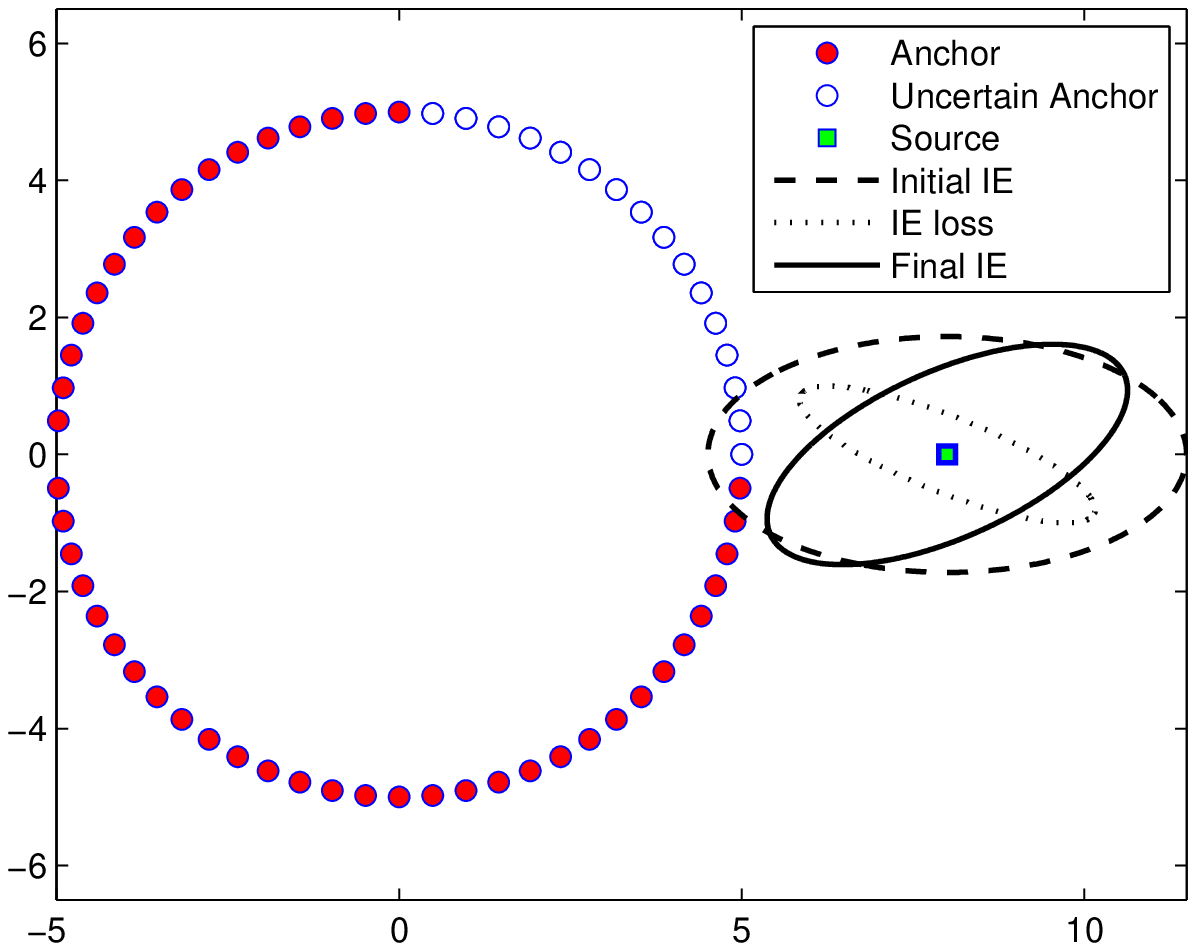}\label{6b}}
\caption{Source location IE with anchor position uncertainty}
\label{Fig6}
\end{minipage}
\vspace{0.00mm}
\begin{minipage}[!htb]{1\linewidth}
\centering
\subfloat[Eccentricity]{\includegraphics[scale=0.45, height=1.8in]{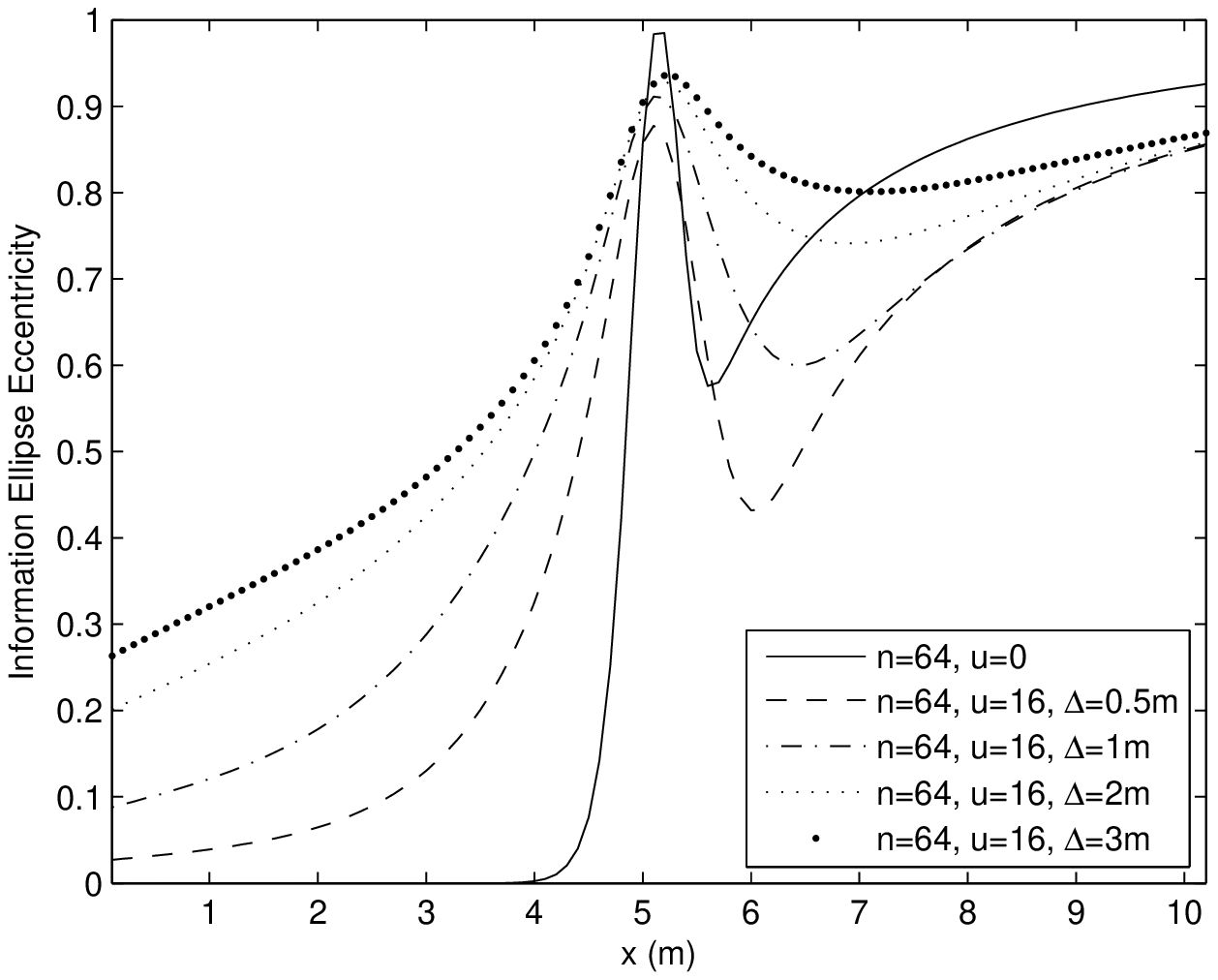}\label{7a}}
\hfil
\subfloat[Area]{\includegraphics[scale=0.45, height=1.8in]{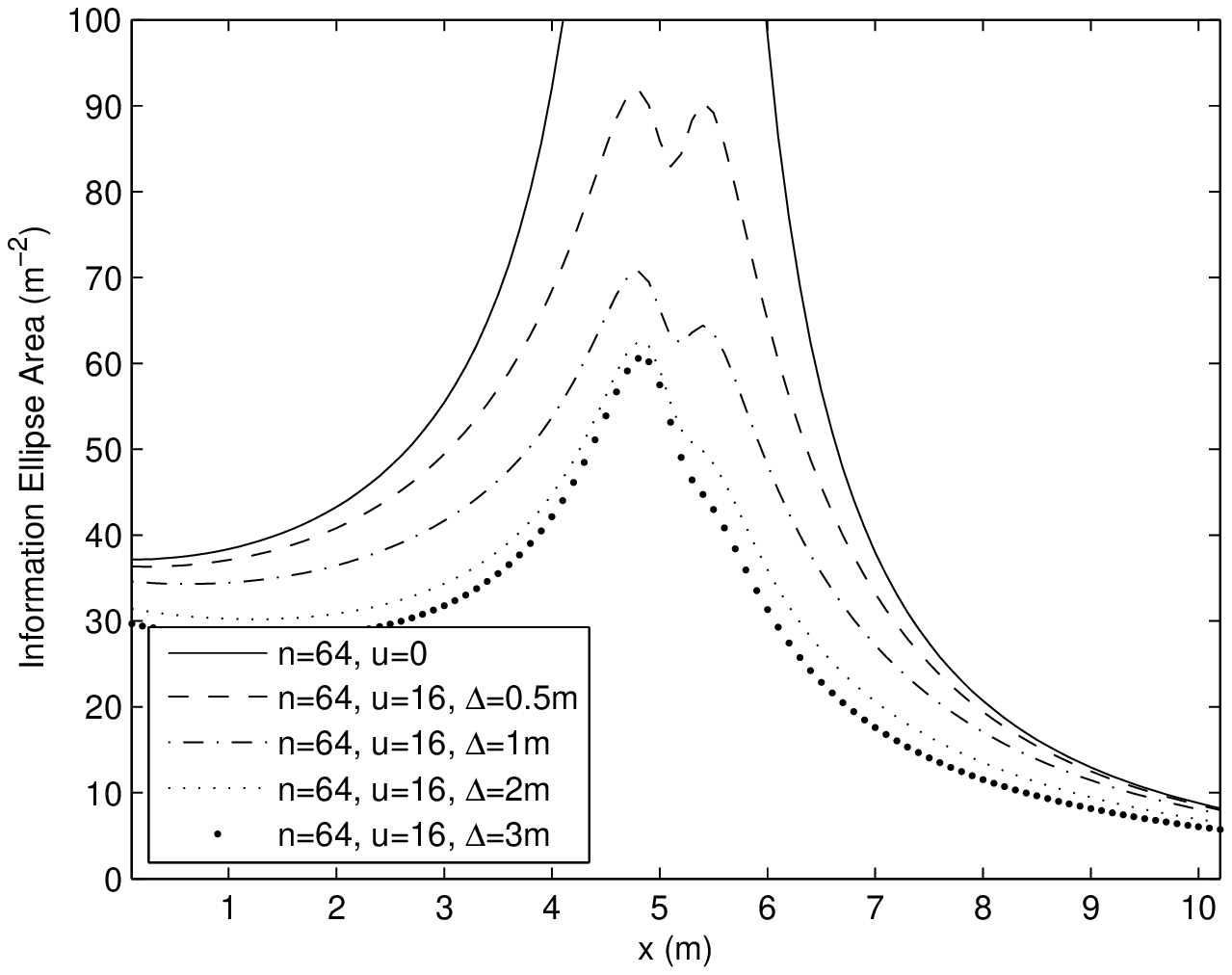}\label{7b}}
\caption{Information Ellipse eccentricity and area}
\label{Fig7}
\end{minipage}
\vspace{0.00mm}
\begin{minipage}[!htb]{1\linewidth}
\centering
\subfloat[$x=4m$]{\includegraphics[scale=0.4, height=1.7in]{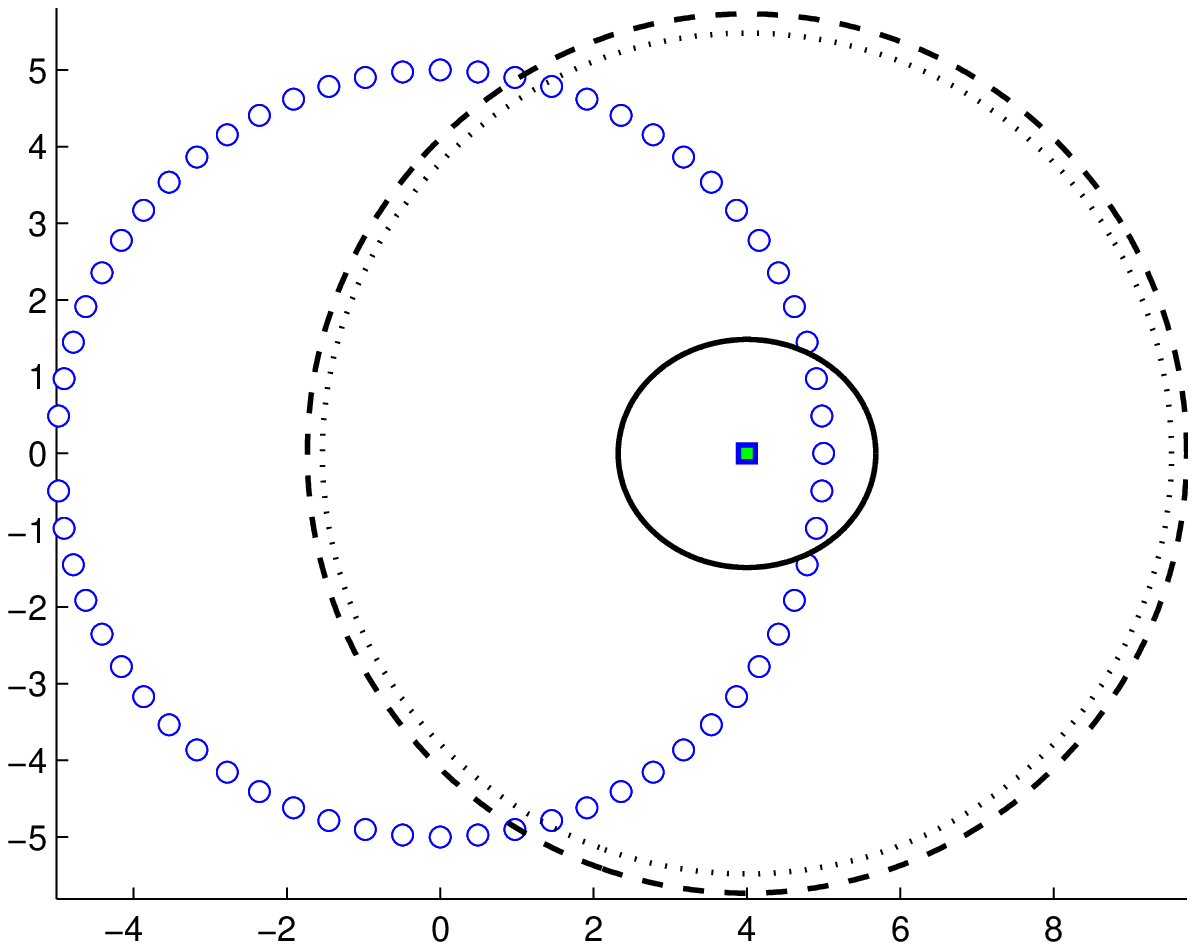}\label{8a}}
\hfil
\subfloat[$x=6m$]{\includegraphics[scale=0.4, height=1.7in]{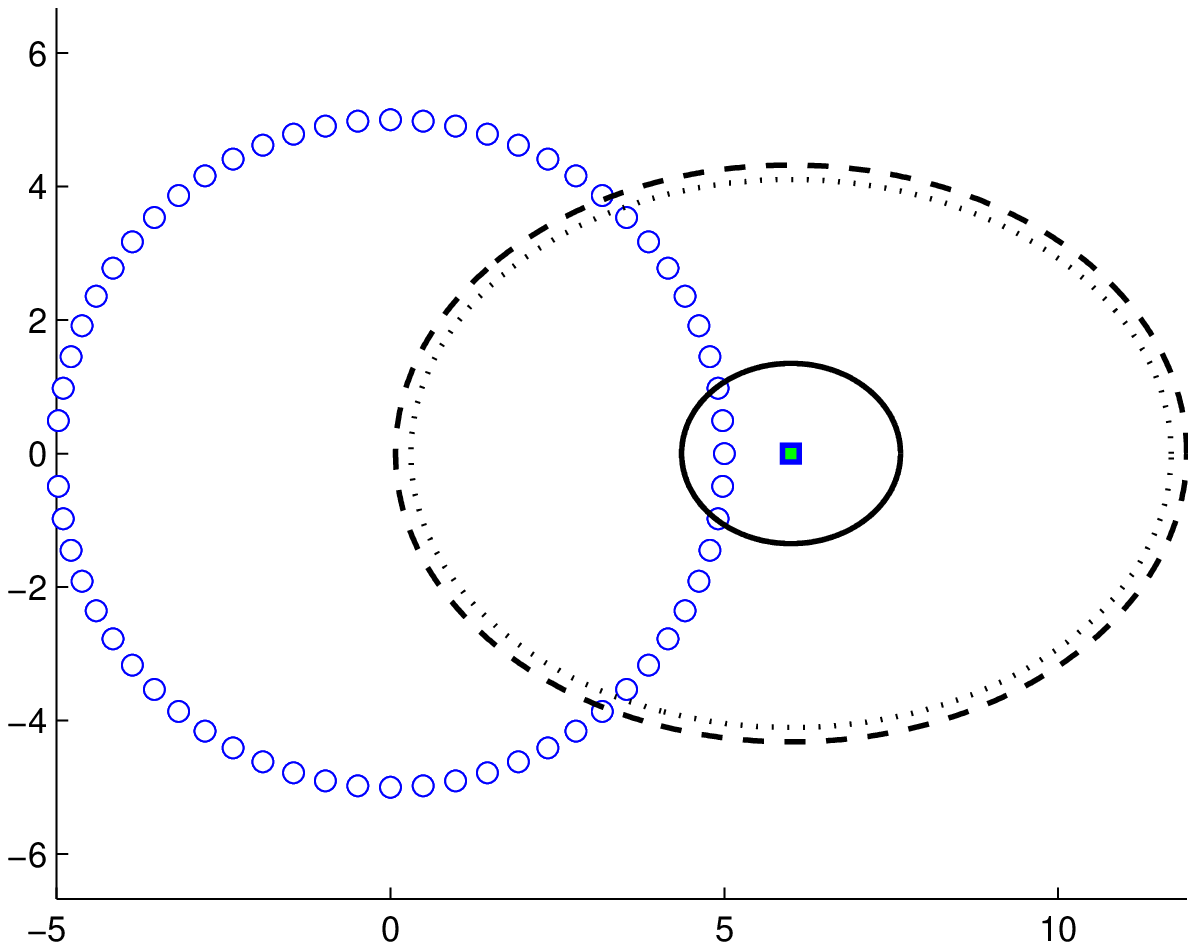}\label{8b}}
\caption{Source location IE with anchor position uncertainty: $U=N$}
\label{Fig8}
\end{minipage}
\end{figure*}

Fig. \ref{Fig6} illustrates the IE changes for different positions of the source both within and outside the circle. Here, $\Delta=3m$. Evidently, the information loss ellipse is directed towards the uncertainty loss source, i.e. towards the first quadrant. The resulting location IE, with smaller area as a result of the anchor uncertainty, is spatially tilted towards the main contributors of source location information. The tilt becomes more evident when the source is outside the circle.

Fig. \ref{Fig7} shows the IE eccentricity and area for different values of $\Delta$. It can be concluded that the increased anchor position uncertainty besides producing IEs with smaller area, it also results in increased eccentricity which is even more evident when the uncertain anchors are clustered in space, i.e. they are non-uniformly distributed.

Fig. \ref{Fig8} illustrates the IE decrease when all $n$ anchors are uncertain. Again, $\Delta=3m$. The FIM for the unknown position of the source is

\begin{equation}
\mathbf{F}(\mathbf{r}_{TX})=\sum\limits_{k=1}^n\frac{\lambda_k}{1+\lambda_k\Delta^2}\mathbf{R}_k.
\end{equation}

The parameters of the IE can be easily calculated using equations (\ref{eq.35.1}), (\ref{eq.35.2}) and (\ref{eq.35.3}). Note that, in this case, since all anchors are positionally uncertain, the source location information IE is severely reduced. However, its spatial orientation remains the same as the initial IE. 

\subsection{Additional transmitting sources: the effect of information gain}

This subsection aims to provide insight into the effect of adding more consecutively active sources on the source location information IE. Equation (\ref{eq.57}) already pointed that several additional location information components appear in this case. The work presented in \cite{22} also illustrates and theoretically shows that activating additional source on known or unknown positions results in improved PEB per active source. For this purpose, this subsection assumes that $8$ additional sources are active (i.e. $S=\left\{1,...,9\right\}$). The sources $2,...,9$ are equidistantly placed on a circle with radii $d_s=1.5d$ around the anchors' circle. The sources' positions are assumed to be known in advance. In such case, the FIM for the unknown position of the source 1, i.e. $\mathbf{r}_{TX(1)}$, obtains the following form ($t=a_k=1,\forall k\in U$)

\begin{equation}
\mathbf{F}(\mathbf{r}_{TX(1)})=\mathbf{\Psi}^{1}-\bigtriangleup\mathbf{\Psi}_1^{1},
\end{equation}

\noindent where $\bigtriangleup\mathbf{\Psi}_1^{1}=\sum\limits_{k\in U}\bigtriangleup\lambda_k^{1}\mathbf{R}_k^1$ and

\begin{equation}
\bigtriangleup\lambda_k^{1}=(\lambda_k^1)^2\Delta^2\big(1-\Delta^2\sum\limits_{p=1}^9(\mathbf{q}_k^1)^T\lambda_k^p(\mathbf{I}_2+\Delta^2\sum\limits_{j=1}^9\lambda_k^j\mathbf{R}_k^j)^{-1}\mathbf{R}_k^p\mathbf{q}_k^1\big).
\label{eq.new2}
\end{equation}

\begin{figure*}[!htb]
\begin{minipage}[!htb]{1\linewidth}
\centering
\subfloat[$x=4m$]{\includegraphics[scale=0.4, height=1.7in]{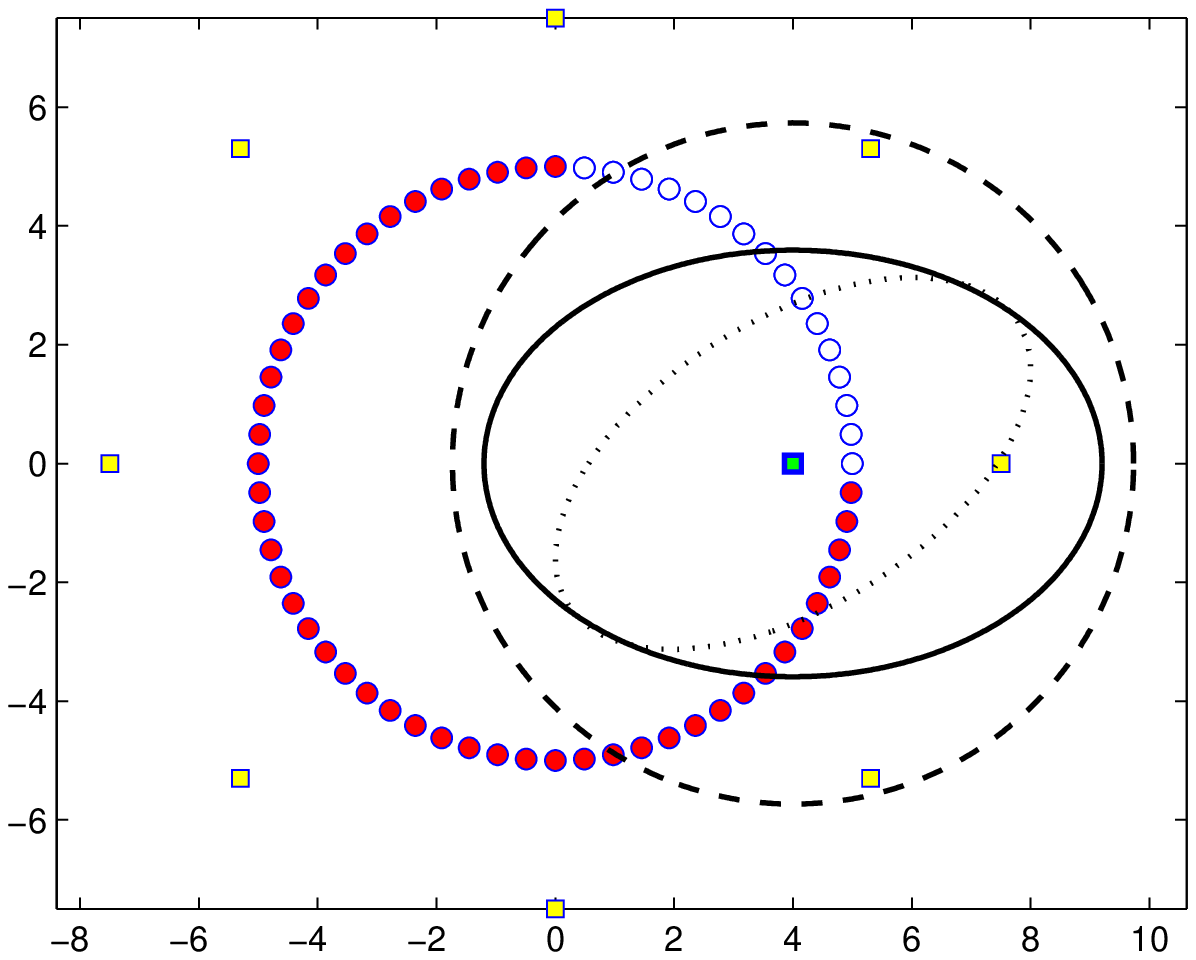}\label{9a}}
\hfil
\subfloat[$x=8m$]{\includegraphics[scale=0.4, height=1.7in]{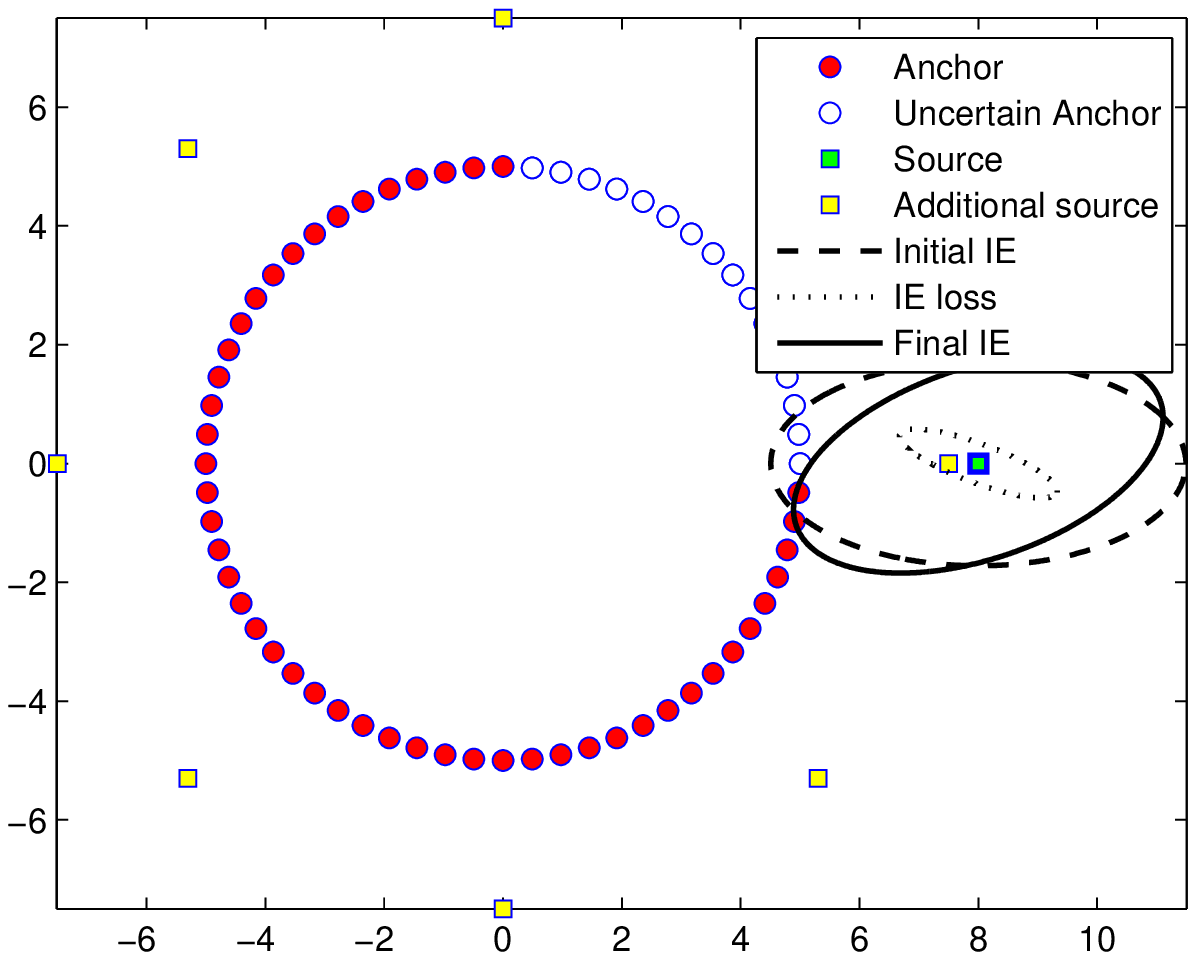}\label{9b}}
\caption{Source location IE with anchor position uncertainty: additional active sources}
\label{Fig9}
\end{minipage}
\vspace{0.00mm}
\begin{minipage}[!htb]{1\linewidth}
\centering
\subfloat[Eccentricity]{\includegraphics[scale=0.45, height=1.8in]{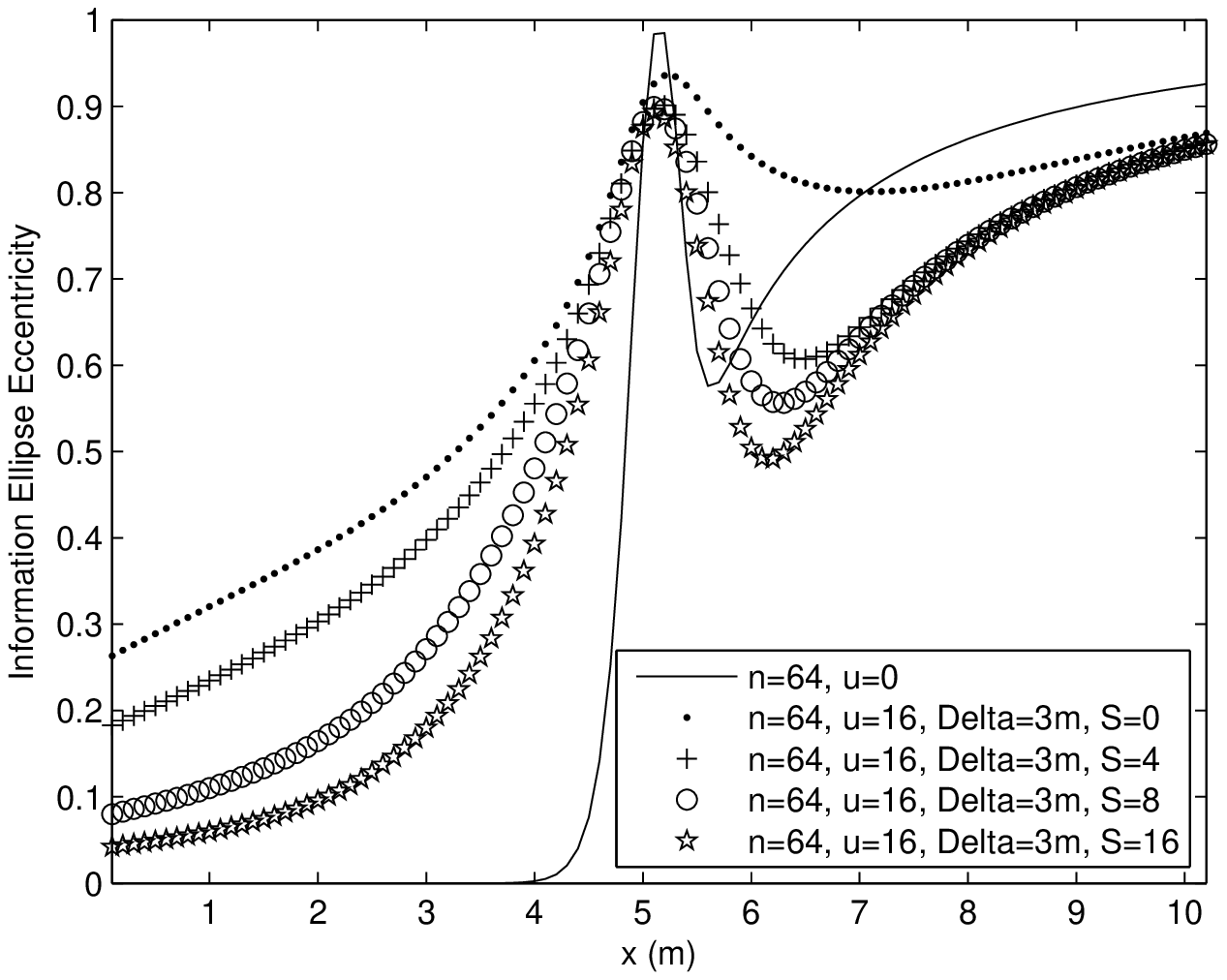}\label{10a}}
\hfil
\subfloat[Area]{\includegraphics[scale=0.45, height=1.8in]{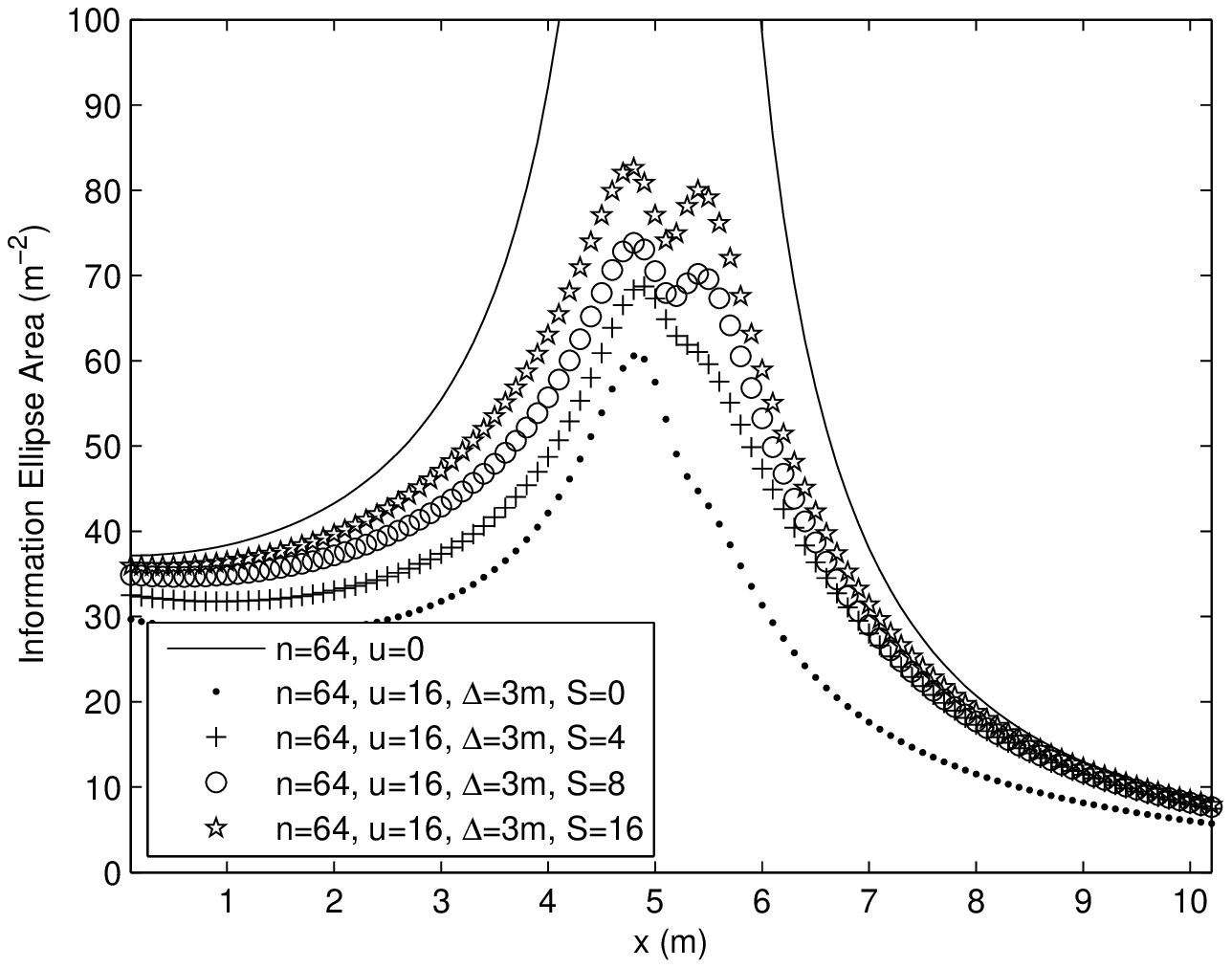}\label{10b}}
\caption{Information Ellipse eccentricity and area: additional active sources}
\label{Fig10}
\end{minipage}
\end{figure*}

Comparing equation (\ref{eq.new2}) with equation (\ref{eq.new1}) it can be concluded that $s$ additional terms appear as a result of activating the sources. These terms decrease the overall reduction $\bigtriangleup\lambda_k^{1}$ per anchor, leading to increased source location information. This is illustrated on Fig. \ref{Fig9}, where it is evident that location information IE increases. Fig. \ref{Fig10} shows the effect of activating additional sources on known positions on the IE eccentricity and area. Note that besides the IE area increase as $s$ increases, the IE eccentricity decreases towards 0, which is desired.

\subsection{Anchor position uncertainty reduction}

Equation (\ref{eq.58}) as well as the work presented in \cite{22} show that arbitrary reduction in the initial anchor position uncertainty (described with the matrix $\mathbf{K}_k$) is possible. In particular, as a result of joint estimation, additional information component (both losses and gains) appear and they contribute towards overall network topology calibration. This subsection focuses on the geometric interpretation of anchor uncertainty reduction due to joint estimation. The initial uncertainty of the $k$-th anchor is given with the following covariance matrix

\begin{equation}
\mathbf{K}_k=
\begin{psmallmatrix}
4 && 1.5 \\
1.5 && 3
\end{psmallmatrix},
\end{equation}

\noindent while the initial anchor position information is quantified with the inverse $\mathbf{K}_k^{-1}$. 

\begin{figure*}[!htb]
\begin{minipage}[!htb]{1\linewidth}
\centering
\subfloat[One source]{\includegraphics[width=2in, height=1.65in]{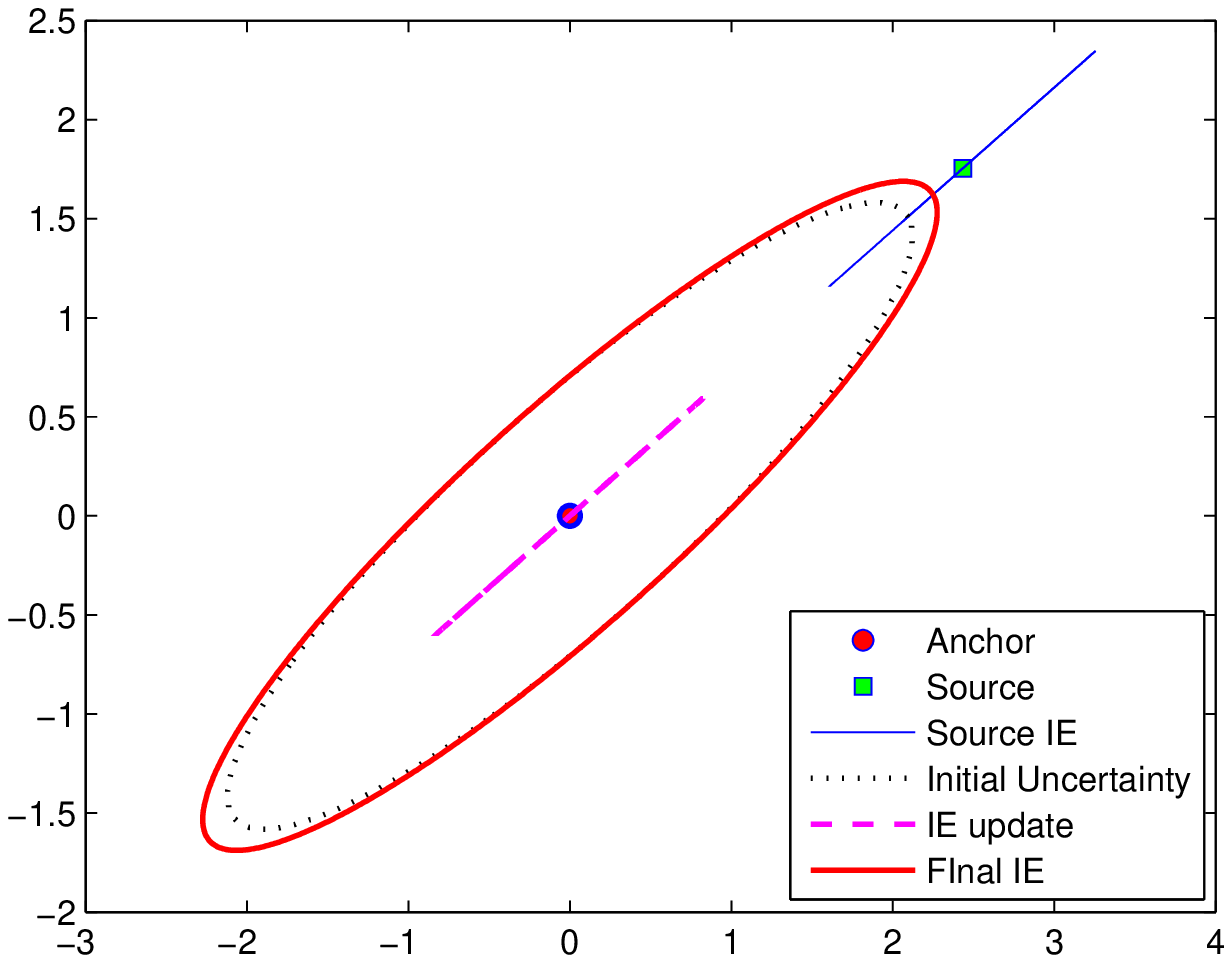}\label{11a}} 
\hfil
\subfloat[Two diagonally aligned sources]{\includegraphics[width=2in, height=1.65in]{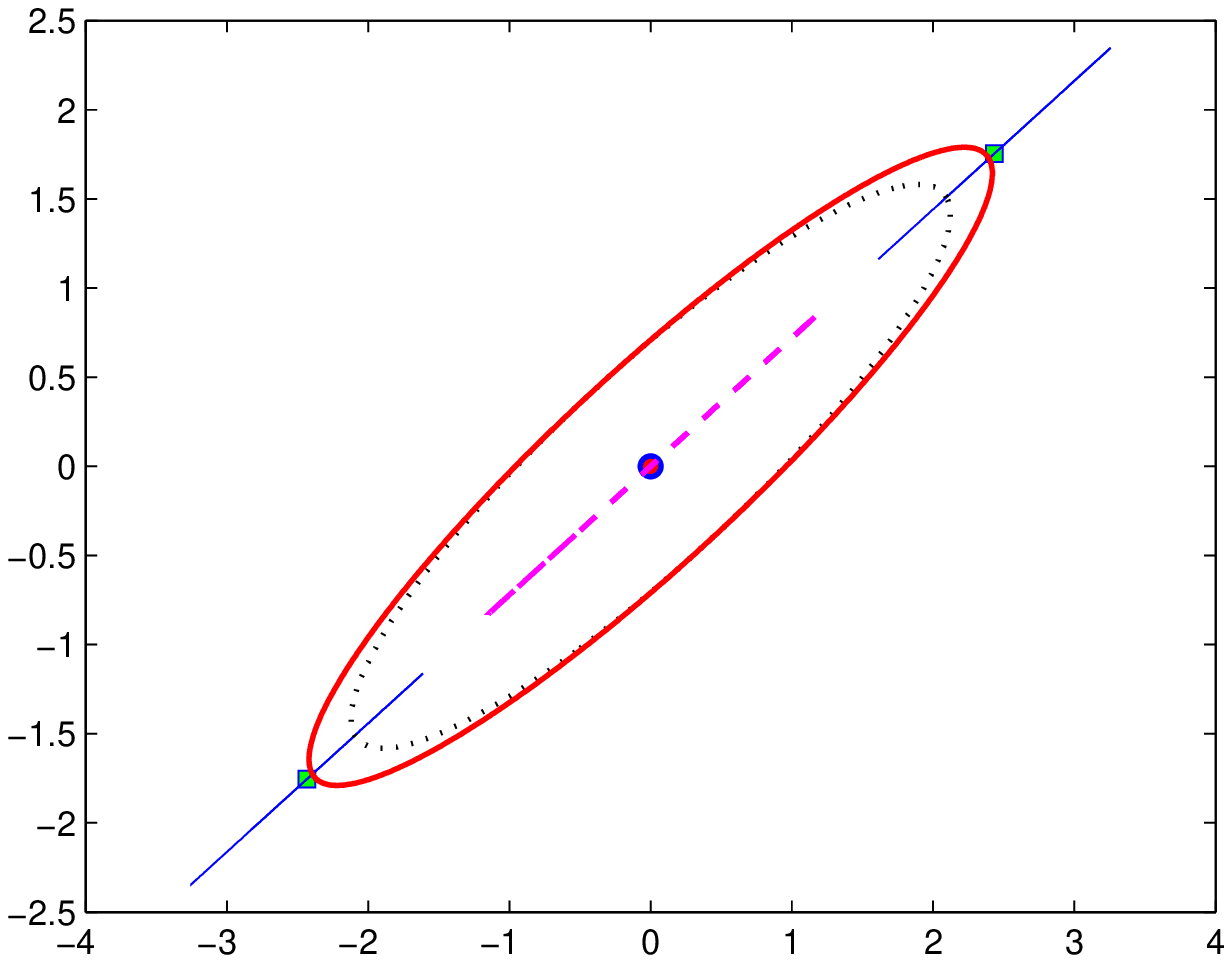}\label{11b}} 
\hfil
\subfloat[Two perpendicular sources]{\includegraphics[width=2in, height=1.65in]{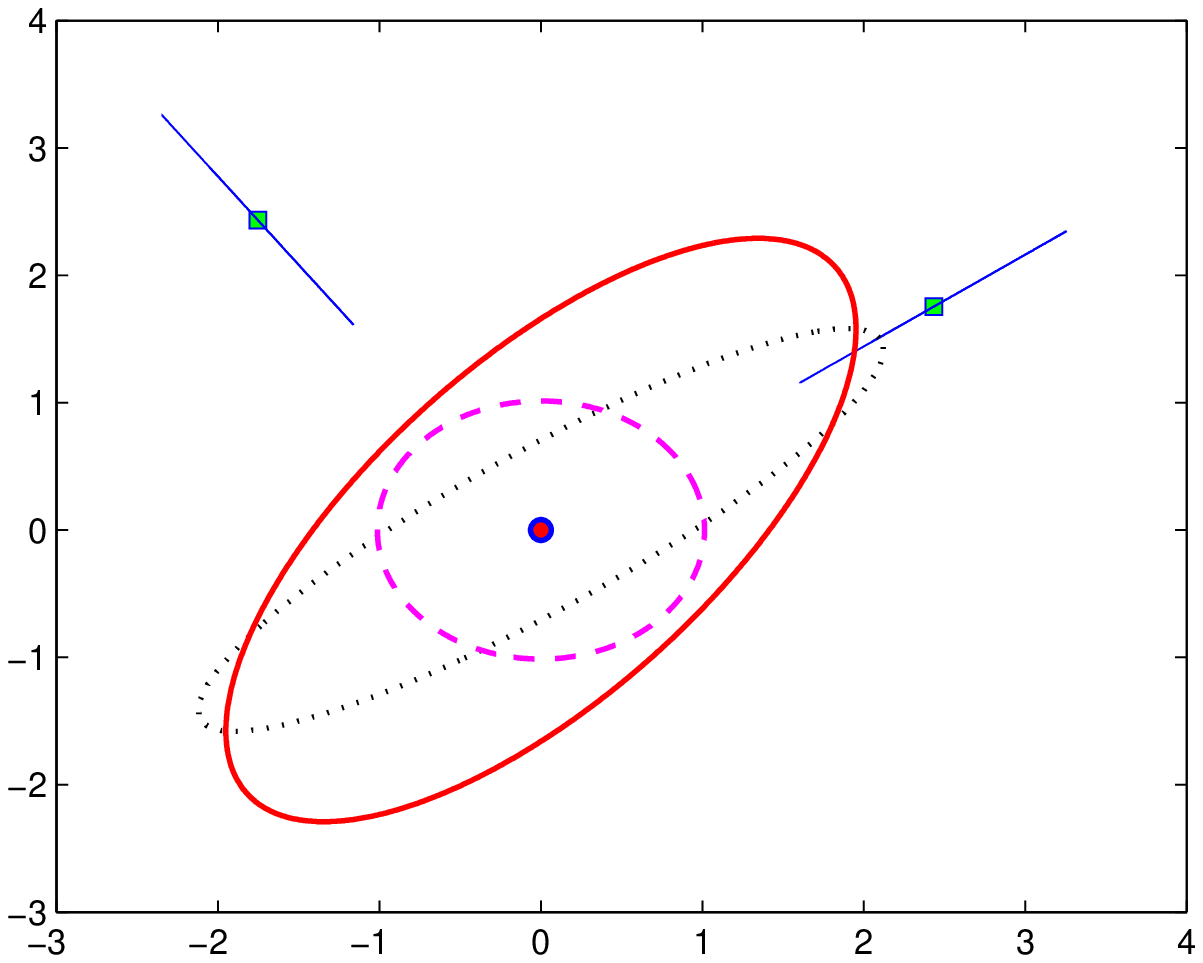}\label{11d}}
\caption{Anchor location information IE updating}
\label{Fig11}
\end{minipage}
\end{figure*}

Fig. \ref{Fig11} illustrates the anchor location IE area increase as a result of the presence of varying number of active transmitters. In this case, the FIM for the unknown position of the anchor is

\begin{equation}
\mathbf{F}(\mathbf{r}_k)_{U=\left\{k\right\}}=\mathbf{K}_k^{-1}+\sum\limits_{j=1}^s\lambda_k^j\mathbf{R}_k^j,
\end{equation}

\noindent i.e., the initial location information about the uncertain anchor $k$ is positively updated $s$ times as a result of the presence of the transmitting sources. Note that this increase, although always positive, will be smaller when the positions of the sources are unknown and jointly estimated with the position of the anchor. In addition, the increase will be even smaller but still positive, if besides the unknown positions of the sources, there are additional anchors with uncertain positions that are jointly estimated.

\begin{figure*}[!htb]
\begin{minipage}[!htb]{1\linewidth}
\centering
\subfloat[Eccentricity]{\includegraphics[scale=0.45, height=1.8in]{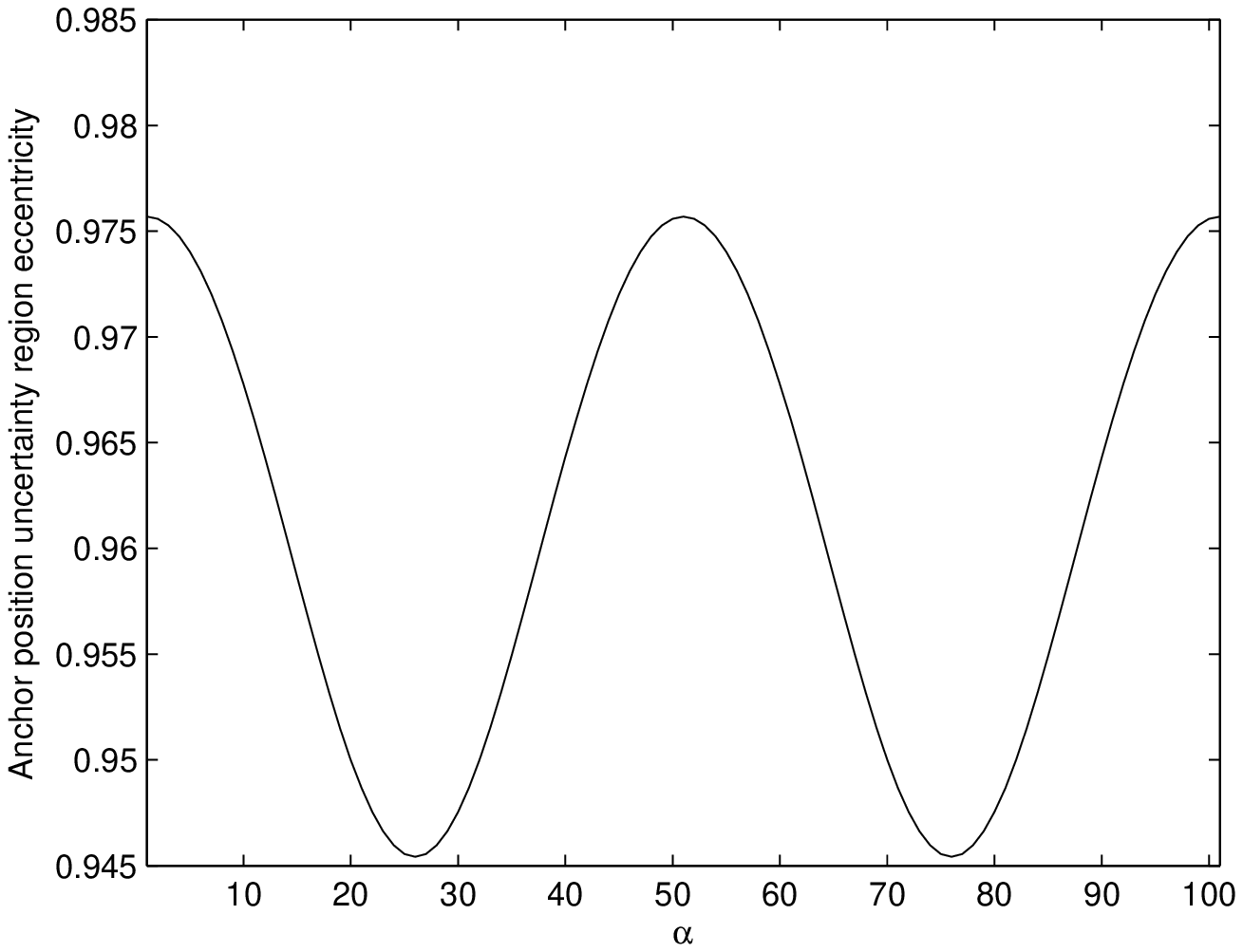}\label{12a}}
\hfil
\subfloat[Area]{\includegraphics[scale=0.45, height=1.8in]{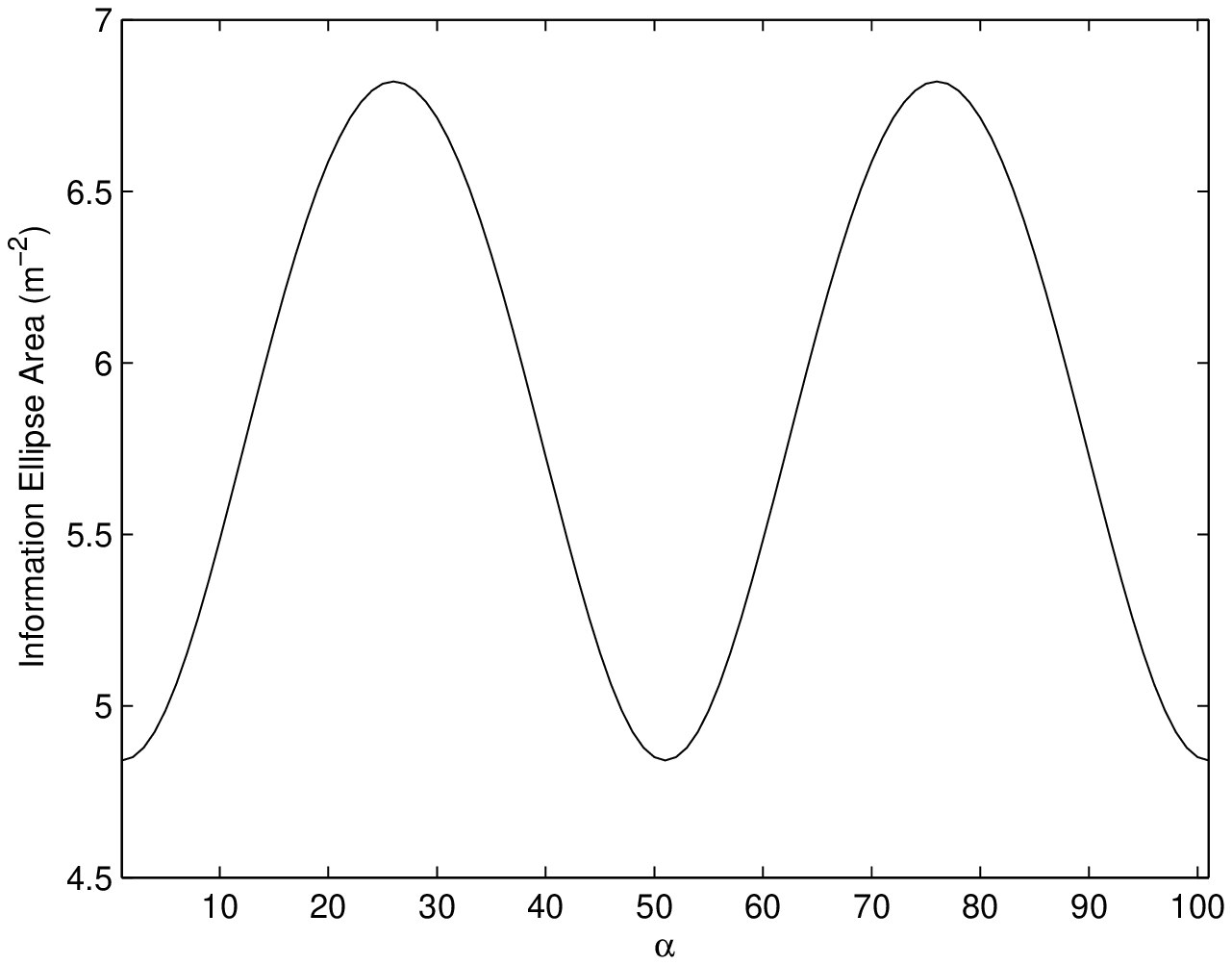}\label{12b}}
\caption{Information Ellipse eccentricity and area: additional active sources}
\label{Fig12}
\end{minipage}
\end{figure*}

\begin{figure*}[!htb]
\begin{minipage}[!htb]{1\linewidth}
\centering
\subfloat[Anchor position IE]{\includegraphics[scale=0.4, height=1.7in]{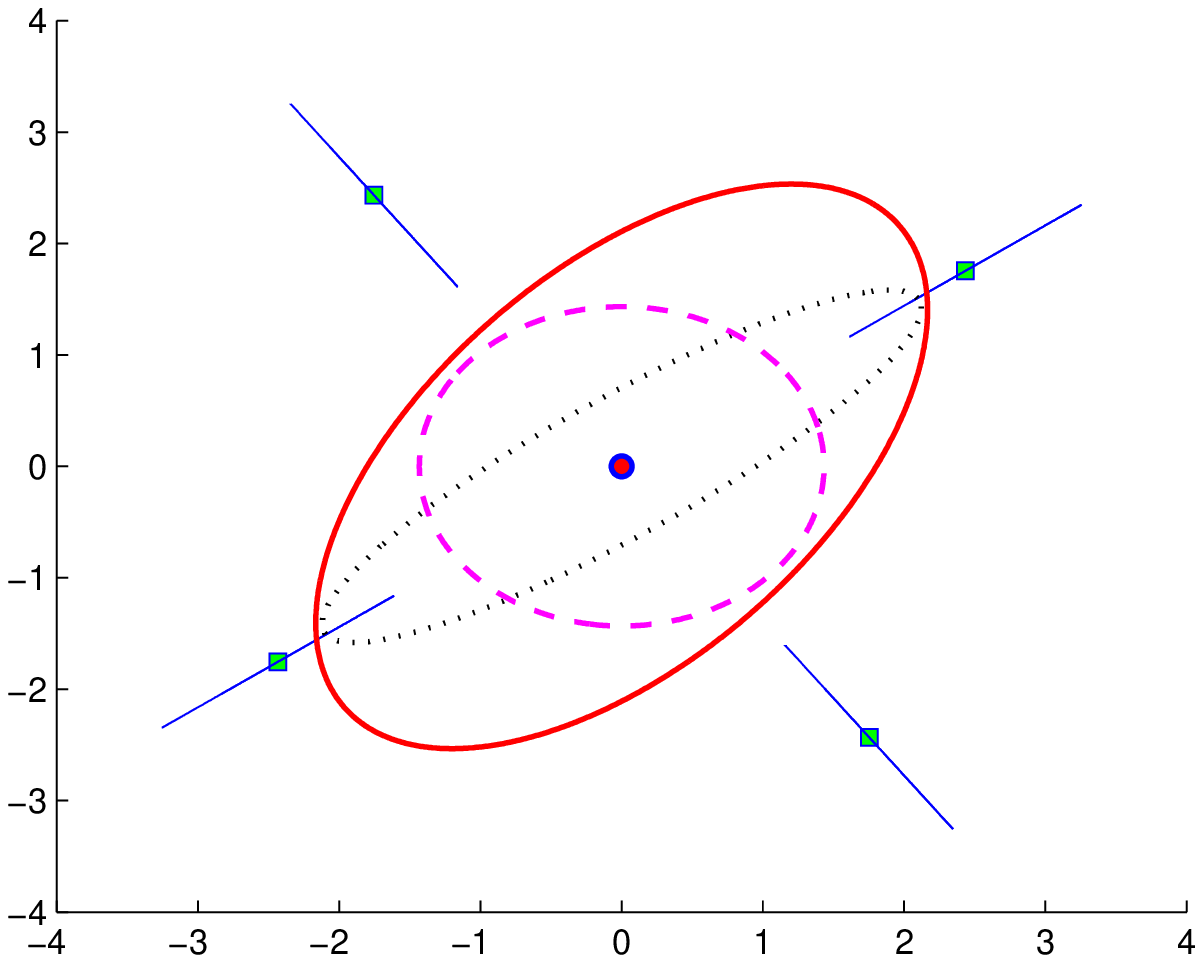}\label{13a}}
\hfil
\subfloat[Anchor position EE]{\includegraphics[scale=0.4, height=1.7in]{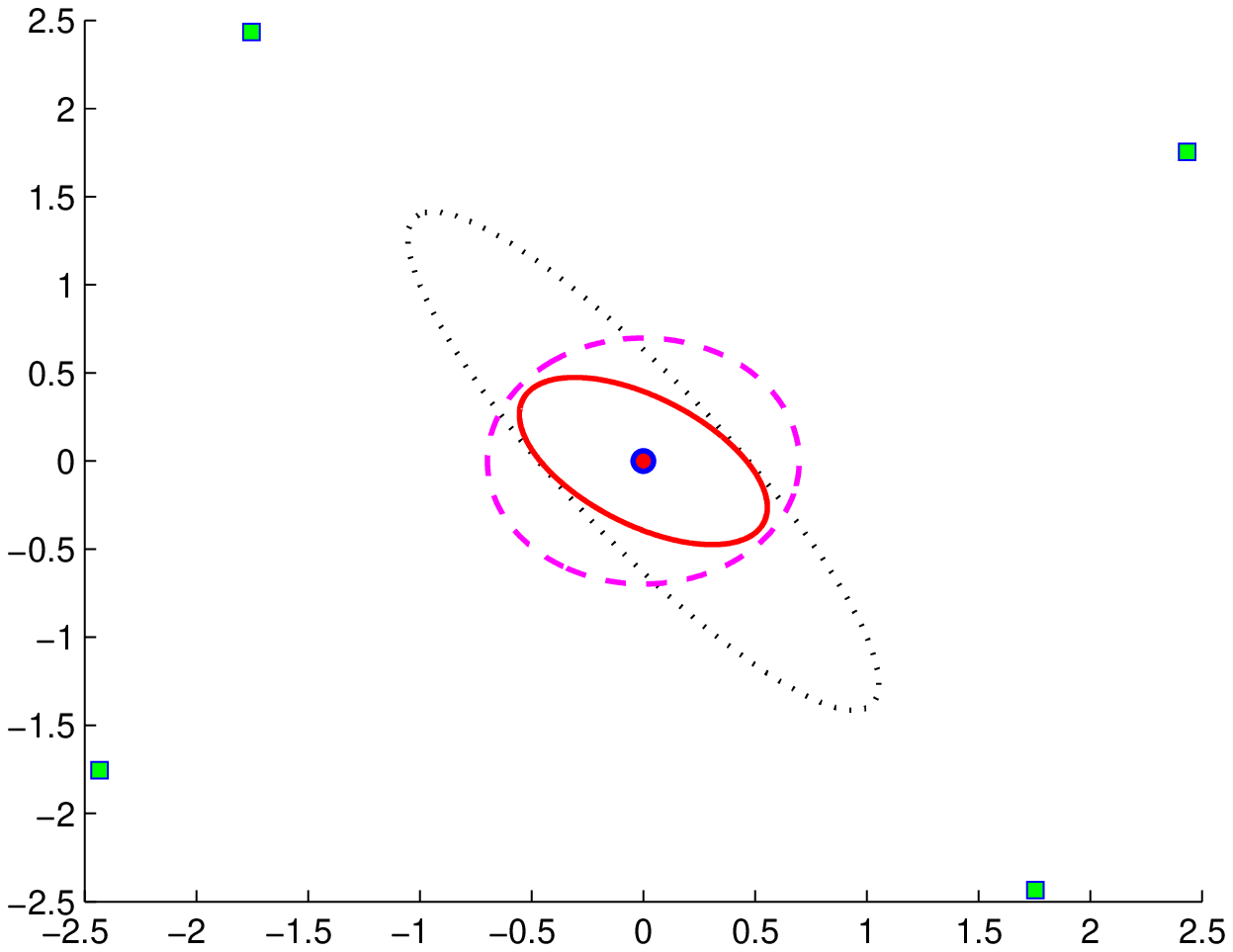}\label{13b}}
\caption{Anchor position Information and Error Ellipses: $s=4$}
\label{Fig13}
\end{minipage}
\end{figure*}

Fig. \ref{Fig12} shows the anchor location information IE eccentricity and area as a function of the angle of the transmitter relative to the angle of the initial anchor location information IE. Note that the eccentricity is minimized and the area maximized when the source is in the direction determined by the smaller semi-axis of the IE. Fig. \ref{Fig13} besides showing the resulting increased IE for the anchor position, also shows the reduced uncertainty region. It can be concluded that, in general, the resulting uncertainty region lies within the intersection of the initial uncertainty region (quantified with $\mathbf{K}_k$) and the sum information gain that stems from the sources.

\section{Conclusions}

This paper presented a geometric interpretation of the theoretical performance bounds of RSS-based source localization in presence of anchor position uncertainty and joint estimation of the unknown sources positions and the unknown positions of the uncertain anchors. The ellipse interpretation is enabled by the positive definiteness and symmetry properties of the Fisher Information Matrix and the Cram{\`e}r-Rao Lower Bound. The benefits and usefulness of the derived geometric representation are illustrated and discussed throughout the paper. The numerical evaluations in the paper provide in-depth analysis of geometrical properties of the information and error bounds with regards to different network configurations and environment parameter settings. Summarizing and concluding, the Information and Error Ellipses provide insights into the geometrical properties of the RSS-based localization in dependence of arbitrary environment and network settings, as well as directions and hints on the various possibilities for designing joint localization strategies that optimally exploit the available location information.

\section*{References}

\bibliography{IEEE_TCOM}

\end{document}